\documentclass[runningheads]{llncs}

\usepackage[utf8]{inputenc}
\usepackage{amsmath, amssymb}
\usepackage{graphicx}
\usepackage{xspace}
\usepackage{tikz,subfigure}
\usetikzlibrary{arrows,shapes,shapes.multipart,snakes,automata,backgrounds,petri,positioning,shadows,matrix,decorations.pathmorphing, decorations.pathreplacing, decorations.markings, fit,positioning,calc,backgrounds,shapes.misc,arrows.meta,fit}
\usepackage{todonotes}

\usepackage[defblank]{paralist}
\usepackage{booktabs}
\usepackage{url}
\usepackage{comment}
\usepackage{stmaryrd}

\usepackage{semantic}
\usepackage{nicefrac}

\usepackage{algorithm}
\usepackage{algorithmic}

\usepackage{thmtools}
\usepackage{thm-restate}

\usepackage{xspace}
\usepackage{wrapfig}
\usepackage{xcolor}

\newcommand{\scheduler}{\mathfrak{S}}
\newcommand{\schedulerT}{\scheduler_{T}}
\newcommand{\schedulerV}{\scheduler_{V}}

\newcommand{\encode}[1]{\underline{#1}}

\newcommand{\lDecode}[2]{#1\llbracket #2 \rrbracket}

\newcommand{\vp}[1]{\encode{\color{red}#1}}
\newcommand{\vd}[1]{\encode{\color{red}#1}}
\newcommand{\vdd}[1]{\encode{\color{red}#1'}{\color{red}}}

\newcommand{\com}[1]{{\color{black!20}\texttt{//}~#1}}

\newcommand{\noP}{\#P}
\newcommand{\noT}{\#T}
\newcommand{\noV}{\#V}

\newcommand{\Runs}[2]{\textsf{Runs}(#1,#2,\G)}
\newcommand{\isGoal}{\textsf{isGoal}}

\newcommand{\mInit}{M_0}
\newcommand{\vInit}{\alpha_0}

\newcommand{\sM}[2]{s_{(#1,#2)}}

\newcommand{\Length}[1]{\mathit{length}(#1)}
\newcommand{\pEnabled}[1]{\pB_{\mathit{enabled}}(#1)}

\newcommand{\pInit}{C_{\mathit{init}}}
\newcommand{\pLoop}{C_{\mathit{loop}}}
\newcommand{\pBody}{GC_{\mathit{body}}}
\newcommand{\pFire}[1]{\pC_{\mathit{fire}}(#1)}
\newcommand{\pIter}[1]{\pLoop^{#1}}

\newcommand{\pDirac}[1]{\delta_{#1}}
\newcommand{\msg}{\mathit{msg}}

\newcommand{\getStep}[1]{\mathit{step}(#1)}

\makeatletter
\g@addto@macro\normalsize{%
\setlength{\abovecaptionskip}{0pt}
\setlength{\belowcaptionskip}{-10pt}
\setlength\abovedisplayskip{3pt}
\setlength\belowdisplayskip{3pt}
\setlength\abovedisplayshortskip{3pt}
\setlength\belowdisplayshortskip{3pt}
}
\makeatother

\newcommand{\E}{\ensuremath{\mathcal{E}}}

\newcommand{\G}{\ensuremath{\mathcal{G}}}

\renewcommand{\S}{\ensuremath{\mathcal{S}}}

\newcommand{\true}{\top}
\newcommand{\set}[1]{\{#1\}}                      

\newcommand{\tup}[1]{\langle #1\rangle}            

\newcommand{\nats}{\ensuremath{\mathbb{N}}\xspace}

\newcommand{\rats}{\mathbb{Q}\xspace}

\newcommand{\bool}{\mathbb{B}\xspace}
\newcommand{\emptysequence}{\ensuremath \epsilon}

\newcommand{\cname}[1]{\ensuremath{\mathsf{#1}}\xspace}

\newcommand{\pFont}[1]{\textnormal{\texttt{#1}}}

\newcommand{\ebnf}{\pFont{::=}}

\newcommand{\pgcl}{\textsc{PPL}}

\newcommand{\pC}{C}

\newcommand{\pCN}{\pC_{sim}}	

\newcommand{\pGC}{GC}
\newcommand{\pB}{B}

\newcommand{\pp}{E}

\newcommand{\pDist}[1]{\mathit{Dist}(#1)}
\newcommand{\pSubDist}[1]{\mathit{SubDist}(#1)}
\newcommand{\pD}{D}
\newcommand{\pMu}{\mu}
\newcommand{\iverson}[1]{[#1]}
\newcommand{\pNormalize}[1]{\mathit{normalize}(#1)}

\newcommand{\pAssign}[2]{#1 ~{\pFont{:=}}~ #2} 

\newcommand{\pObserve}[1]{\pFont{observe}~#1}
\newcommand{\pCompose}[2]{#1;#2}
\newcommand{\pDo}{\pFont{do}}
\newcommand{\pOd}{\pFont{od}}
\newcommand{\pIf}{\pFont{if}}
\newcommand{\pFi}{\pFont{fi}}
\newcommand{\pIffi}[1]{\pIf~#1~\pFi}
\newcommand{\pDood}[1]{\pDo~#1~\pOd}
\newcommand{\pDump}{\pFont{log}}

\newcommand{\pChoice}[1]{~\pFont{[]}~}
\newcommand{\pGuard}[3]{#1 ~\xrightarrow{#2}~ #3}
\newcommand{\pTab}{\qquad}

\newcommand{\pState}{s}
\newcommand{\pLog}{\ell}
\newcommand{\sUpdate}[2]{[#1 \mapsto #2]}

\newcommand{\pVar}{\textnormal{\textbf{Var}}}
\newcommand{\pStates}{\textnormal{\textbf{PS}}}
\newcommand{\pLogs}{\textnormal{\textbf{PL}}}
\newcommand{\psl}{\textnormal{\textbf{PSL}}}

\newcommand{\pDone}[1]{\textnormal{\textsf{guards}}(#1)}
\newcommand{\pBranchProb}[1]{\textnormal{\textsf{weight}}(#1)}

\newcommand{\pSem}[1]{\llbracket #1 \rrbracket}
\newcommand{\pSSem}[1]{\mathcal{S}\llbracket #1 \rrbracket}

\newcommand{\ie}{i.e.\xspace}

\newcommand{\net}{N}

\newcommand{\pre}[1]{{{}^\bullet{#1}}} 
\newcommand{\post}[1]{{{#1}^\bullet}} 
\newcommand{\goto}[1]{\ensuremath[{#1}\rangle}
\newcommand{\preco}{\textnormal{\textsf{pre}}}
\newcommand{\postco}{\textnormal{\textsf{post}}}

\newcommand{\nEnabled}[1]{\textnormal{\textsf{frontier}}(#1)}

\definecolor{deepblue}{HTML}{0C3B80}
\definecolor{deepgreen}{HTML}{2EA601}
\definecolor{lightOrange}{HTML}{FFA03C}
\definecolor{darkOrange}{HTML}{F1800A}
\definecolor{lightBlue}{HTML}{0174CD}
\definecolor{greenF}{HTML}{2CBB5C}
\definecolor{cyan}{HTML}{86A6D5}
\definecolor{darkred}{HTML}{8B0000}

\newcommand{\plname}[1]{\mathtt{#1}}
\newcommand{\trname}[1]{\mathtt{#1}}
\tikzstyle{place}=[circle,thick,draw=black,fill=white,minimum size=7mm,font=\fontsize{9}{144}\selectfont]
\tikzstyle{transition}=[rectangle,thick,draw=black,fill=gray!20,minimum size=7mm]

\usepackage[capitalise]{cleveref}

\makeatletter
\newcommand{\customlabel}[2]{%
\protected@write \@auxout {}{\string \newlabel {#1}{{#2}{}}}}
\makeatother

\title{Data Petri Nets meet \\ Probabilistic Programming (Extended version)}
 \author{%
 Martin Kuhn \inst{1} \and
 Joscha Gr\"uger \inst{1,2} \and
 Christoph Matheja \inst{3} \and
 Andrey Rivkin\inst{3} 
 }

 \institute{
     German Research Center for Artificial Intelligence (DFKI), SDS Branch Trier, Trier, Germany
     \email{martin.kuhn@dfki.de}
     \and
     University of Trier, Germany \email{grueger@uni-trier.de}
     \and
     Technical University of Denmark, Kgs. Lyngby, Denmark
     \email{\{chmat,ariv\}@dtu.dk}
 }

\begin{document}

\maketitle

\begin{abstract}
Probabilistic programming (PP) is a programming paradigm that allows for writing statistical models like ordinary programs, performing simulations by running those programs, and analyzing and refining their statistical behavior using powerful inference engines. This paper takes a step towards leveraging PP for reasoning about data-aware processes. To this end, we present a systematic translation of Data Petri Nets (DPNs) into a model written in a PP language whose features are supported by most PP systems. We show that our translation is sound and provides statistical guarantees for simulating DPNs. 
Furthermore, we discuss how PP can be used for process mining tasks and report on a prototype implementation of our translation. We also discuss further analysis scenarios that could be easily approached based on the proposed translation and available PP tools.
\end{abstract}

\section{Introduction}
\label{sec:introduction}

Data Petri nets (DPNs)~\cite{MannhardtLRA16,LeoniFM18} is a popular formalism for data-aware processes that is used in business process management (BPM) and process mining (PM) for various tasks including discovery~\cite{MannhardtLRA16}, conformance checking~\cite{MannhardtLRA16,FelliGMRW21,FelliGMRW22}, formal verification and correctness analysis~\cite{FelliMW22,FelliMW22-ijcar}.
Moreover, it has been shown in~\cite{LeoniFM21} that DPNs can formalize the integration of a meaningful subset of BPMN with DMN S-FEEL decision tables.
Recent work also addresses stochastic~\cite{MannhardtLSL23} and uncertainty-related~\cite{FelliGMRW22} aspects of DPNs.

\smallskip
\noindent\emph{Simulation for DPNs.}
One of the key techniques in the BPM and PM repertoires is \emph{simulation}, which allows for flexible analyses, such as ``what-if'' analysis,
that often cannot be addressed by formal verification
or that touch upon non-functional aspects (e.g., time, costs, resources)
that are not reflected in process models~\cite{Aalst15,Aalst18}.
For DPNs, the prime application of simulation is the generation of synthetic event logs aiming at closing the gap between missing datasets needed for substantial evaluation of discovery and analysis techniques. 
To the best of our knowledge, \cite{Grueger21} is the only approach 
that explicitly provides a DPN simulation engine which is grounded in DPN semantics and allows to perform randomised generation of fixed-length executions. 

\smallskip
\noindent\emph{Probabilistic Programming (PP)}~\cite{gordon2014,webPPL14}
is a paradigm developed by the programming languages and machine learning communities to make statistical models and reasoning about them with Bayesian inference accessible to non-experts.
The key idea is to represent statistical models as programs and leave the development of efficient simulation and inference engines to the language developers.

A probabilistic program can be thought of as an ordinary program with the ability to sample from probability distributions.
Running such a program means performing stochastic simulation: a single program execution corresponds to a single simulation of the underlying model. 
Modern PP systems have two characteristic features: First, they support \emph{conditioning} the possible executions (or \emph{feasible simulations}) on observed evidence, e.g. to refine a synthetic model using real-world data or user knowledge.
Second, they support \emph{inference techniques} to compute or approximate the \emph{probability distribution} modeled by a program.

This raises the question whether one can leverage the existing PP machinery for simulating DPNs instead of developing ad-hoc simulators.

\smallskip
\noindent\emph{Contributions and outline.}
In this paper, we explore whether and how probabilistic programs are a suitable abstraction for simulating DPNs such that (1) the simulation process is based on a statistical model clearly defined as a probabilistic program and (2) simulation, event log generation, and further (statistical) analyses could benefit from using PP together with its sampling and inference capabilities (cf.~\cite{gordon2014,webPPL14}). 
Our main contributions can be summarized as follows:
\begin{compactitem}[$\bullet$]
	\item We formalize an execution semantics for DPNs with schedulers, which are used in discrete-event simulation to resolve non-determinism~\cite{Law15} ($\rightarrow$ \textbf{Section~\ref{sec:dpn}}).
	\item We formalize the essence of many existing probabilistic programming languages ($\rightarrow \textbf{Section~\ref{sec:ppl}}$) and then develop a novel systematic encoding of DPNs with schedulers into a probabilistic programming language ($\rightarrow$ \textbf{Section~\ref{sec:encoding}}).
	\item We show that our encoding is correct, \ie our PP encoding of DPN produces exactly the runs of the  encoded DPNs and preserves the probabilities of all simulated runs with respect to the scheduler ($\rightarrow$ \textbf{Section~\ref{sec:correcntess}}).
	\item We discuss how to leverage our encoding and inference engines provided by PP systems for various Process Mining tasks. ($\rightarrow$ \textbf{\Cref{sec:tasks}}).
	\item We report on a proof-of-concept implementation of our encoding into the PP language webPPL~\cite{webPPL14} and discuss two case studies ($\rightarrow$ \textbf{\Cref{sec:conclusions}}).
\end{compactitem}

\smallskip
\noindent\emph{Related work.}
There are multiple approaches for generating synthetic logs via model simulation for data-aware processes that can potentially be applied to DPNs. 
Given the well-known relation between colored Petri nets~\cite{cpnbook} and DPNs~\cite{LeoniFM18}, one may use CPN Tools\footnote{\url{https://cpntools.org/}} to produce logs without noise or incompleteness~\cite{Medeiros2004} or logs in which effects of workload on processing speeds are recorded~\cite{Nakatumba2012}.
To the best of our knowledge, CPN Tools is the only applicable (non-commercial) tool in which one can explicitly define schedulers for simulation tasks.

Alternatively,
one may resort to approaches that generate multi-perspective logs for BPMN 2.0 models (and rely on the connection between BPMN and DPNs established in~\cite{LeoniFM21}).
For example, in~\cite{Mitsyuk2017} the authors rely on an executable BPMN semantics 
and do not only support standard elements (e.g., gateways, sub-processes), but also 
data objects for driving exclusive choices to generate random (multi-perspective) event longs. 
Similarly,~\cite{Burattin2015} generates multi-perspective logs by running a randomized play-out game.

To the best of our knowledge,~\cite{Grueger21} is the 
simulation engine that explicitly targets DPNs.
It is based on a direct implementation of the execution semantics of DPNs that randomly fires enabled transitions. The random choices cannot be directly influenced, e.g. by changing the underlying scheduler.
Moreover, any statistical guarantees about the generated event logs provided by the engine -- e.g., are the generated traces representative with respect to an underlying stochastic process? -- are, at best, ad hoc. The same holds for the above works on the randomised log generation for BPMN.

One may also rely on studied relationships between business process models and discrete event simulation (DES)~\cite{RosenthalTS21}. For example,~\cite{PufahlWW17} applies DES to BPMN 2.0 models to produce multi-perspective logs. 
While our work is grounded in Bayesian statistics, DES takes a frequentist approach to simulation where all ``variables'' are already known by the domain expert. 
We refer to~\cite{Chick06} for a discussion of the advantages of Bayesian approaches compared to DES.

\section{Preliminaries}
\label{sec:components}
\noindent\textbf{Sequences.}
A finite \emph{sequence} $\sigma$ over a set $S$ of length $|\sigma| = n \in \nats$ is a function $\sigma : \{1,\ldots,n\} \to S$. 
We denote by $\emptysequence$ the empty sequence of length $n=0$.
If $n > 0$ and $\sigma(i) = s_i$, for $1\leq i \leq n$, we write $\sigma = s_1\ldots s_n$. 
The set of all finite sequences over $S$ is denoted by $S^{*}$. 
The \emph{concatenation} $\sigma = \sigma_1 \sigma_2$ of two sequences $\sigma_1,\sigma_2 \in S^{*}$ is given by $\sigma : \{ 1, \ldots, |\sigma_1|+|\sigma_2|\}\rightarrow S$, such that $\sigma(i) = \sigma_1(i)$ for $1 \leq i \leq |\sigma_1|$, and $\sigma(i) = \sigma_2(i - |\sigma_1|)$ for $|\sigma_1|+1 \leq i \leq |\sigma_1|+|\sigma_2|$.

\smallskip
\noindent\textbf{Probability Distributions.}
A (discrete)
\footnote{For simplicity, we will work with discrete distributions. However, many PP systems also support continuous distributions.}
\emph{subprobability distribution} over a non-empty, countable set $X$ is a function $\pMu\colon X \to [0,1]$ s.t. $\sum_{x \in X} \mu(x) \leq 1$.
We say that $\pMu(x)$ is the probability assigned to $x \in X$ and call $\pMu$ a \emph{distribution} if $\sum_{x \in X} \pMu(x) = 1$.
We denote by $\pDist{X}$ (resp. $\pSubDist{X}$) the set of all (sub)distributions over $X$, respectively.
We consider a few examples:
\begin{compactenum}
	\item The \emph{Dirac distribution} $\delta_{x}\colon X \to [0,1]$ assigns probability $1$ to a fixed element $x \in X$ and probability $0$ to every other element of $X$.
	\item For two distinct elements $x,y \in X$ and some $p \in [0,1]$, the \emph{Bernoulli distribution} $B(p,x,y)\colon X \to [0,1]$ models a coin-flip with bias $p$ and possible outcomes $x$ and $y$: it assigns $p$ to $x$, $(1-p)$ to $y$, and $0$ otherwise. 
	\item The \emph{uniform distribution} $\pFont{unif(a,b)}\colon \rats \to [0,1]$ assigns probability $\nicefrac{1}{(b-a+1)}$ to values in $[a \,..\, b]$; to every other rational number $r \in \rats$, it assigns $0$.
	\item The function $\nicefrac{B}{2}\colon \rats \to [0,1]$ given by $\nicefrac{B}{2}(0) = \nicefrac{B}{2}(1) = \nicefrac{1}{4}$ and $\nicefrac{B}{2}(x) = 0$ for all $x \in \rats \setminus \{0,1\}$ is a subdistribution in $\pSubDist{\rats}$, which represents the Bernoulli distribution $B(\nicefrac{1}{2},0,1)$ scaled by $\nicefrac{1}{2}$.
\end{compactenum}
For a subdistribution $\pMu \in \pSubDist{X}$, its \emph{normalized distribution} is given by
$
  \pNormalize{\pMu} ~=~ \frac{\pMu}{\sum_{x \in X} \pMu(x)}
$. 
For example, normalizing $\nicefrac{B}{2}$ yields the distribution
$
  \pNormalize{\nicefrac{B}{2}} ~=~ \frac{\nicefrac{B}{2}}{\nicefrac{1}{2}} ~=~ B(\nicefrac{1}{2},0,1)$.
As in standard probability theory, we assume $\nicefrac{0}{0} = 0$, \ie if 
$\pMu$ assigns probability $0$ everywhere, so does $\pNormalize{\pMu}$.

\section{Data Petri Nets with Probabilistic Schedulers}
\label{sec:dpn}

Data Petri nets (DPNs)~\cite{MannhardtLRA16,LeoniFM18} extend traditional place-transition nets with the possibility to manipulate scalar case variables, which are used 
to constrain the evolution of the process through 
\emph{guards} assigned to transitions. 
Each guard is split into a pre- and postcondition that is defined over two variable sets, $V$ and $V'$, where $V$ is the set of case variables and $V'$ keeps their primed copies used to describe variable updates.\footnote{Hereinafter, we assume that $X'$ defines a copy of the set $X$ in which each element $x\in X$ is primed. The same holds for individual elements.} 
We denote by $\E(X,\Delta)$ the set of all \emph{boolean} expressions over variables in $X$ and constants in $\Delta$. 
For an expression $e$, $V(e)$ and $V'(e)$ denote the sets of all unprimed and primed variables in $e$, respectively.

\begin{definition}[Data Petri net]
A \emph{data Petri net} (DPN) is a tuple
$\net = \tup{P, T, F, l, A, V, \Delta, \preco,\postco}$, where:
\begin{inparaenum}[\it (i)]
	\item $P$ and $T$ are finite, disjoint sets of \emph{places} and \emph{transitions}, respectively;
	\item $F\subseteq(P \times T)\cup(T \times P)$ is a \emph{flow relation};
	\item $l:T\rightarrow A$ is a \emph{labeling} function assigning
 activity names from $A$ to every transition $t\in T$;
	\item $V$ is a set of \emph{case variables} and $\Delta$ is a \emph{data domain};
        \item $\preco:T\to\E(V,\Delta)$ is a transition \emph{precondition-assignment} function; and
        \item $\postco:T\to\E(V\cup V',\Delta)$ is a transition \emph{postcondition-assignment} function.
\end{inparaenum}
\end{definition}
Given a DPN $\net = \tup{P, T, F, l, A, V, \Delta, \preco,\postco}$, 
we will from now on write $P_{\net}$, $T_{\net}$, etc. to denote $\net$'s components; we omit the subscript if the given net is clear from the context. 
Given a place or transition $x\in (P_N \cup T_N)$ of $N$, the \emph{preset} $\pre{x}$ and the
\emph{postset} $\post{x}$ are given by
$\pre{x}=\set{y\mid (y,x)\in F}$ and
$\post{x}:=\set{y\mid (x,y)\in F}$.

We next turn to the DPN execution semantics. 

\begin{definition}[DPN state]
A \emph{state} of a DPN $\net$ is a pair $(M,\alpha)$, where
\begin{inparaenum}[\it(i)]
	\item $M:P_N\to\mathbb{N}$ is a total \emph{marking} function, assigning a number $M(p)$ of \emph{tokens}
		to every place $p\in P_N$ and
	\item $\alpha:V_N\to\Delta$ is a total \emph{variable valuation} function
		assigning a value to every variable in $V_N$.
\end{inparaenum}
\end{definition}
A DPN moves between states by firing (enabled) transitions.
 Every state, together with a (variable) valuation $\beta$ inducing an update over (some of) the net variables,
 yields a set of \emph{enabled} transitions,
 which can be \emph{fired} to progress the net. 
 After a transition fires, a new state
 is reached, with a new corresponding marking and valuation.

\begin{definition}[Transition enabling, firing]
\label{def:dpn-semantics}
Let $\net$ be a DPN, $(M,\alpha)$ a state, $t\in T_N$ a transition, and $\beta:(V_N\cup V_N')\nrightarrow\Delta_N$ be a partial valuation function.

We denote by $(M,\alpha)[(t,\beta)\rangle$ that the \emph{step} $(t,\beta)$ is \emph{enabled} in $(M,\alpha)$, i.e.
\begin{compactenum}[\it(i)]
	\item\label{cond:enabled-1} 
 $\beta$ is defined for all the variables in $\preco(t)$ and $\postco(t)$ and no other variables; 
	\item\label{cond:enabled-2} for every $v\in V_N(\preco_N(t)\land\postco_N(t))$, we have that $\beta(v) = \alpha(v)$, i.e.,
		$\beta$ matches $\alpha$ on all the (non-primed) variables of $t$ that are not being modified;
        \item\label{cond:enabled-3}  for every $p \in \pre{t}$, we have $M(p) \geq 1$;
        \item\label{cond:enabled-4}  $\beta\models \preco_N(t)\land \postco_N(t)$, i.e., $\beta$ satisfies the pre- and post-conditions of $t$.
\end{compactenum}
Moreover, we denote by $(M, \alpha) \goto{(t,\beta)} (M', \alpha')$ that the state $(M', \alpha')$ is the result of \emph{firing} transition $t$ according to step $(t,\beta)$, i.e.
$(M,\alpha)[(t,\beta)\rangle$ holds and 
\begin{inparaenum}[\it (1)]	
		\item for every $p\in P_N$, we have $M'(p) = M(p) - F_N(p,t) + F_N(t,p)$;
		\item  $\alpha'(v)=\beta(v')$ for all $v'\in V'_N(\postco(t))$;  and
            \item $\alpha'(v)=\alpha(v)$ for all $v\in V_N(\preco_N(t)\land\postco_N(t))$.
	\end{inparaenum}

\end{definition}

A \emph{run} of a DPN $\net$ is a sequence
of steps $\sigma=(t_1, \beta_1)\ldots (t_{k}, \beta_{k})$, where $k\in\nats$.
A state $(M',\alpha')$ is \emph{reachable} from $(M,\alpha)$,
if there is a run $\sigma$ as above such that 
$(M,\alpha)\goto{(t_1, \beta_1)}(M_1,\alpha_1)\goto{(t_2, \beta_2)}\cdots \goto{(t_k, \beta_k)} (M', \alpha')$,
denoted $(M,\alpha)\goto{\sigma}(M', \alpha')$.
We denote by $\S_{\net}$ the set of all states of $\net$. 
A run $\sigma$ is \emph{legal} iff there are two states $(M,\alpha),(M',\alpha')\in \S_{\net}$ s.t. 
$(M,\alpha)\goto{\sigma}(M',\alpha')$. 

\smallskip
\noindent \textbf{Simulation with schedulers.}
Simulating a DPN corresponds to generating (some of its)  runs. 
A crucial part of the simulation process is how to resolve non-deterministic choices attributed to multiple, simultaneously 
enabled transitions and possibly infinitely many partial valuations the transitions' pre- and post-conditions. 
This issue has been resolved in discrete event system simulation using schedulers that handle non-deterministic choices~\cite{Law15}.
In the same vein, we define randomized schedulers for DPNs.

\begin{definition}[DPN Scheduler]\label{def:scheduler}
	A \emph{scheduler} of a DPN $\net$ is a function
	\begin{align*}
		\scheduler\colon \S_{\net} \to \pDist{T_{\net}} \times (V_{\net}' \to \pDist{\Delta_{\net}}).
	\end{align*}
\end{definition}
In words, a scheduler $\scheduler$ assigns to every state $(M,\alpha)$ a probability distribution $\scheduler_T$ over the net's transitions and a mapping $\scheduler_V$ from the net's (primed) variables to probability distributions over domain values.
Intuitively, $\scheduler_T$ resolves the nondeterminism that arises if multiple transitions are enabled: if all transitions are enabled, the probability of picking transition $t$ is given by $\scheduler_{T}(t)$; if some transitions are not enabled, their probability is uniformly distributed amongst the enabled transitions.
The function $\scheduler_V$ resolves for every $v' \in V_{\net}'$ the nondeterminism that arises from postconditions that hold for different evaluations of $v'$: the probability of assigning $\cname{d} \in \Delta_{\net}$ to $v'$ is given by $\scheduler_V(v')(\cname{d})$. We will modify this probability by conditioning on the fact that a transition's postcondition must hold after firing it.
Notice that postconditions may introduce dependencies between case variables, even though their values are sampled independently.

\begin{figure}[t!]
    \centering
	\resizebox{.43\textwidth}{!}{
\begin{tikzpicture}[->,>=stealth',auto,x=25mm,y=1.0cm,node distance=16mm 	and 16mm,thick]
			
\node[place,label=left:$\plname{p_0}$,xshift=2mm] (p0) at (0,0) {};

\node[transition,below of = p0, yshift=5mm,label=right:$\trname{init}$] (init) {};
\node[left of = init,xshift=-4mm]  (g1)  {$\postco:\,{t'>\cname{0}\land o'=\cname{0}}$};
\node[place,below of = init, xshift=-10mm,label=right:$\plname{p_1}$] (p1) {};

\node[transition,left of = p1,label=below:$\trname{bid}$] (bid) {};
\node[above of = bid,yshift=-9mm]  (g2)  {$\begin{array}{@{}c@{}} {\preco:\, t>0}\\{\postco:\,o'>o} \end{array} $};
\node[place,below of = init, xshift=10mm,label=left:$\plname{p_2}$] (p2) {};

\node[transition,right of = p2,label=below:$\trname{timer}$] (timer) {};
\node[above of = timer,yshift=-9mm]  (g3)  {$\begin{array}{@{}c@{}} {\preco:\, t>0}\\{\postco:\,t'<t} \end{array} $};

\node[transition,below of = p1,xshift=10mm,label=left:$\trname{hammer}$] (hammer) {};
\node[right of = hammer,xshift=2mm]  (g5)  {$\preco:\, t\leq\cname{0}\land o>\cname{0} $};

\node[place,below of = hammer,yshift=5mm,label=below:$\plname{p_3}$] (p3) {};

                    
\node[transition,right of = p3,xshift=10mm,label=below:$\trname{reset}$] (reset) {};
\node[above of = reset,yshift=-11mm]  (g4)  {$\preco:\, o=\cname{0}$};

					\path[]
   				 (p0) edge (init)
   				 (init) edge (p1)
   				 (init) edge (p2)
   				 (p1) edge [bend left=15] (bid)
   				 (bid) edge [bend left=15] (p1)
   				 (p2) edge [bend left=15] (timer)
   				 (timer) edge [bend left=15] (p2)
   				 (p1) edge (hammer)
   				 (p2) edge (hammer)
   				 (hammer) edge (p3)
   				 (p3) edge (reset)
    			;
       \draw[->,rounded corners=5pt] (reset.east) -- ($(reset.east)+(8mm,0)$) |- (p0.east);
\end{tikzpicture}
}
  	\caption{A simple auction process~\cite{FelliMW22}}
  	\label{fig:dpn-ex}
\end{figure}
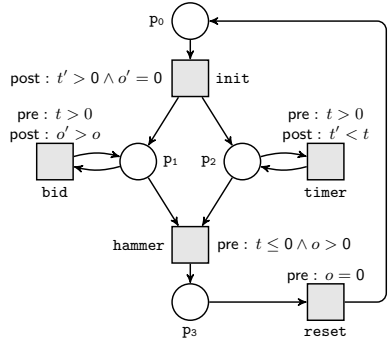
\noindent
\emph{Example 1.}\customlabel{ex1}{1}
Consider the DPN $N$ in Figure~\ref{fig:dpn-ex} representing a simple auction process, where the last offer is stored in $o$ and the time progression is captured by $t$. 
One possible scheduler uniformly selects an enabled transition, \ie $\scheduler_T^{auc} = \pFont{unif}(1,|T_{N}|)$. 
Alternatively, one can model priorities between potentially simultaneously enabled transitions by selecting a different distribution that depends, e.g. on the value of the timer.
Examples of distributions for selecting values for the variables $t'$ and $o'$, include the Poisson (over countable sets of timestamps and outcomes), uniform (over finite subsets of timestamps and outcomes such as $\scheduler^{auc}_V(t') = \pFont{unif}(0,99)$ and $\scheduler^{auc}_V(o') = \pFont{unif}(1,30)$), geometric, and normal distribution.\footnote{To improve readability, our approach is formalized for discrete distributions. However, continuous distributions are supported by many PP systems.} In the following examples, we will use $\scheduler^{auc}=(\scheduler_T^{auc},\scheduler_V^{auc})$ as the default scheduler. 
\hfill{\footnotesize$\triangle$}

We will now quantify the probability of a run in terms of the likelihood\footnote{We use ``likelihood'' to refer to an ``unnormalized probability'', ie. a value that is obtained from a subdistribution. The sum of all likelihoods may thus be less than 1.} of each involved step.
To this end, let 
$\nEnabled{M,\alpha} ~=~ \{ t ~|~ \exists \beta\colon (M,\alpha)[(t,\beta)\rangle\}$ be the set of transitions that are enabled in a given state.
Moreover, given a Boolean proposition $p$, we denote by $\iverson{p}$ its indicator function, \ie $\iverson{p} = 1$ if $p$ is $\true$ (true), and $0$ if $p$ is $\bot$ (false).

\begin{definition}[Step Likelihood]\label{def:step-likelihood}
Given a net $\net$ and a scheduler $\scheduler$ with $\scheduler(M,\alpha) = (\schedulerT,\schedulerV)$, the \emph{likelihood} of a step $(t,\beta)$
  in a state $(M,\alpha)$ is
 \begin{align*}
  	\mathbb{P}_{\scheduler}[M,\alpha]\left( t,\beta \right) 
  	~=~ 
  	\underbrace{\mathbb{P}_{\scheduler}[M,\alpha]\left( t \right)}_{\textnormal{likelihood of selecting $t$}}
  	\,\cdot\,
  	\underbrace{\mathbb{P}_{\scheduler}[M,\alpha]\left( \beta ~|~ t \right)}_{\textnormal{likelihood of selecting $\beta$ given $t$ }},
  \end{align*}
  where the likelihood of selecting transition $t$ among the enabled ones is
  \begin{align*}
    \mathbb{P}_{\scheduler}[M,\alpha]\left( t \right)
    \quad=\quad &
    \frac{\iverson{t\in\nEnabled{M,\alpha}}\cdot\schedulerT(t)}{\sum\limits_{t' \in T_{\net}} \iverson{t' \in \nEnabled{M,\alpha}} \cdot \schedulerT(t')}
  \end{align*}
  and the likelihood of selecting $\beta\colon (V_N\cup V'_N)\nrightarrow\Delta_N$ 
  given transition $t$ s.t. $\beta(v) = \alpha(v)$, for every $v\in V_N(\preco_N(t)\land\postco_N(t))$, 
  is
  \begin{align*}
    \mathbb{P}_{\scheduler}[M,\alpha]\left( \beta ~|~ t \right) 
    \quad=\quad &
    \iverson{\beta\models \postco_N(t)} \cdot
    \prod_{x'\in V'_N(\beta)} \scheduler_V(x')(\beta(x')).
  \end{align*}
\end{definition}
In the spirit of Bayes' rule, the likelihood
of a step $(t,\beta)$ is the likelihood that the scheduler selects transition $t$ multiplied with the likelihood of selecting valuation $\beta$ given the previous selection of $t$. Notice that $\mathbb{P}_{\scheduler}[M,\alpha]\left( t,\beta \right)$ is $0$ if the chosen transition $t$ is not enabled -- and thus $\iverson{t\in\nEnabled{M,\alpha}} = 0$ -- or if the chosen $\beta$ does not satisfy $t$'s postcondition -- and thus $\iverson{\beta\models \postco_N(t)} = 0$.

\noindent
\emph{Example 2.}\customlabel{ex2}{2}
Consider the DPN and scheduler 
from Ex.~\ref{ex1}. The likelihood of the step $(\trname{timer},\beta)$, where $\beta=\set{t\mapsto 20,o\mapsto 5,t'\mapsto 11}$, in state $(M,\alpha)$, where $M(\plname{p_1})=M(\plname{p_2})=1$ and $M(\plname{p_0})=M(\plname{p_3})=0$ and $\alpha=\set{t \mapsto 20,o \mapsto  5}$, is $\mathbb{P}_{\scheduler^{auc}}[M,\alpha]\left( \trname{timer},\beta \right)=\frac{\nicefrac{1}{5}}{\nicefrac{3}{5}} \cdot \nicefrac{1}{100}$.
\hfill{\footnotesize$\triangle$}

We expand the above likelihood measure to DPN runs that reach a possibly infinite set $\G$ of \emph{goal states} for the first time. Examples of goal states include all states with a final marking or all states in which $x\geq 15$.
To this end, $\Runs{M}{\alpha}$ denotes the set of all runs $\sigma = (t_1,\beta_1)\ldots (t_n,\beta_n)$, for all $n \geq 0$, such that, for all $k \in \{1, \ldots, n-1\} $, 
\begin{align*}
	\text{if}\quad (M,\alpha) \goto{(t_1,\beta_1)\ldots (t_k,\beta_k)} (M_k,\alpha_k) \quad\text{then} \quad (M_k,\alpha_k) \notin \G.
\end{align*}
Notice that not all runs in $\Runs{M}{\alpha}$ are legal and only some \emph{may} reach states from $\G$. 
We then define the likelihood that a run $\sigma \in \Runs{M}{\alpha}$ reaches $\G$ from $(M,\alpha)$ via its step likelihoods, or $0$ if no goal state is ever reached.

\begin{definition}[Likelihood of a Run]\label{def:run-likelihood}
  Given a DPN $\net$, a scheduler $\scheduler$, a state $(M,\alpha)$, goal states $\G$, and a run $\sigma \in \Runs{M}{\alpha}$ over $\net$,
  the \emph{likelihood} $\mathbb{P}_{\scheduler}[M,\alpha]\left( \sigma \models \diamondsuit \mathcal{G} \right)$\footnote{As in \cite{mc-book}, we use an LTL-like notation to denote that $\sigma$ ends in a state in $\G$.} that $\sigma$ reaches $\G$ from $(M,\alpha)$ is defined recursively as
  \begin{align*}
 	\begin{cases}
 		1, 
 		& \text{if } (M,\alpha) \in \mathcal{G} \\
 		\mathbb{P}_{\scheduler}[M,\alpha]\left( t,\beta \right)
 		\cdot
 		\mathbb{P}_{\scheduler}[M',\alpha']\left( \sigma' \models \diamondsuit \mathcal{G} \right),
 		& \text{if } \sigma = (t,\beta) \, \sigma' \text{ and } (M,\alpha) \goto{(t,\beta)} (M',\alpha') \\
 		0, & \text{otherwise}.
 	\end{cases}
 \end{align*}
\end{definition}
Technically, the likelihood of a run is defined analogously to the reachability probability of a trace in an (infinite-state) Markov chain -- a well-established stochastic model for describing sequences of events (cf.~\cite{puterman,mc-book}).
The probability of the next step may thus depend on the \emph{current} state but is independent of previously visted states.
While this assumption is common and allows for dependencies between cases, e.g. case variables can be accessed and modified by different transitions, it can be seen as a limitation because those dependencies must be modeled explicitly in the DPN's state space.

\noindent
\emph{Example 3.}\customlabel{ex3}{3}
Consider the DPN from Ex.~\ref{ex1}, set of goal states $\G=\set{(M,\alpha)\in \S_N\mid M(\plname{p_3})=1}$ and a run 
$\sigma=(\trname{init},\set{t'\mapsto 10})
(\trname{bid},\set{t\mapsto 10,o'\mapsto 5})
(\trname{timer},\set{t\mapsto 10,t'\mapsto 0})
(\trname{hammer},\set{t\mapsto 0,o\mapsto 5})$.
Then the likelihood to reach one of the goal states from initial state $(M_0,\alpha_0)=(\set{\plname{p_0}\mapsto 1,\plname{p_1},\plname{p_2},\plname{p_3}\mapsto 0},\set{t\mapsto 0,o\mapsto 0})$ is  
$\mathbb{P}_{\scheduler^{auc}}[M_0,\alpha_0]\left( \sigma \models\diamondsuit \mathcal{G} \right)=
(1 \cdot \nicefrac{1}{100}) \cdot 
(1 \cdot \nicefrac{1}{10}) \cdot 
(\nicefrac{1}{2} \cdot \nicefrac{1}{100}) \cdot 
(\nicefrac{1}{3} \cdot 1)
$.
\hfill{\footnotesize$\triangle$}

To obtain the \emph{probability} of a run, we then normalize the above likelihood with the likelihoods of all runs that can only reach $\G$ after firing all of their steps. 

\begin{definition}[Probability of a Run]\label{def:run-probability}
    Given a DPN $\net$, a scheduler $\scheduler$, a state $(M,\alpha)$, goal states $\G$ and a run $\sigma$ over $\net$,
    the \emph{probability that $\sigma$ reaches $\G$} 
    given that all runs starting in $(M,\alpha)$ eventually reach $\G$ is defined as
   \begin{align*}
   	\mathbb{P}_{\scheduler}[M,\alpha]\left( \sigma ~|~ \diamondsuit \mathcal{G} \right)
   	~=~
   	\frac{\mathbb{P}_{\scheduler}[M,\alpha]\left( \sigma \models \diamondsuit \mathcal{G} \right)}{\sum_{\sigma' \in \textsf{Runs}(M,\alpha,\G)}\mathbb{P}_{\scheduler}[M,\alpha]\left( \sigma' \models \diamondsuit \mathcal{G}\right)}.
 \end{align*}
 \end{definition}

Notice that a run $\sigma$ is legal if $\mathbb{P}_{\scheduler}[M,\alpha]\left( \sigma ~|~ \diamondsuit \mathcal{G} \right)>0$ holds,
because the likelihood is set to $0$ if we attempt to fire a transition that is not enabled or select a valuation that does not satisfy the transition's postcondition.
Moreover, one can easily extract sequences of events, i.e. traces from a legal run and its initial state.

Our translation to probabilistic programs in \Cref{sec:encoding} will make sure 
that, given a scheduler $\scheduler$ and a set of goal states $\G$, every run $\sigma$ is produced with the probability $\mathbb{P}_{\scheduler}[M,\alpha]\left( \sigma ~|~ \diamondsuit \mathcal{G} \right)$.

\section{The Probabilistic Programming Language $\pgcl$}
\label{sec:ppl}

In this section, we give the necessary background on probabilistic programming by introducing a small probabilistic programming language called $\pgcl$ whose features are supported by many existing PP systems (cf.~\cite{gordon2014,webPPL14}).

\begin{figure}
    \centering
  \begin{minipage}{.575\textwidth}
\begin{align*}
     \pC ~\ebnf~ &
     \pAssign{x}{\pD} \tag{probabilistic assignment}
     \\ & ~|~
     \pObserve{\pB} \tag{condition on event $\pB$}
     \\ & ~|~ 
     \pDump~\msg \tag{add $\msg$ to the log}
     \\ & ~|~ 
     \pCompose{\pC}{\pC} \tag{sequential composition}
     \\ & ~|~
     \pDood{\pGC} \tag{unbounded loop}
     \\ & ~|~
     \pIffi{\pGC} \tag{conditional}
     \\
     \pGC ~\ebnf~ & 
     \pGuard{\pB}{\pp}{\pC} \tag{guarded command} 
     \\ & ~|~
     \pGC\,\pChoice{}\,\pGC \tag{choice}
  \end{align*}
    \caption{Syntax of $\pgcl$.}
    \label{fig:pgcl-syntax}
    \end{minipage}%
    \hspace{1cm}
    \begin{minipage}{.3\textwidth}
  {\small\begin{align*}
  	& \pAssign{x}{\pFont{uniform}(1,3);} \\
  	& \pIf \\
  	& \qquad \pGuard{x = 1}{\nicefrac{1}{3}}{\pAssign{y}{4}} \\
  	& \pChoice{} ~ \pGuard{x > 1}{\nicefrac{1}{3}}{\pAssign{y}{5}} \\
  	& \pChoice{} ~ \pGuard{x > 1}{\nicefrac{1}{3}}{\pAssign{y}{x + 2}} \\
  	& \pFi; \\
  	& \pObserve{{\color{black}\pB}}
  \end{align*}}%
   \caption{Example program.}
    \label{fig:pgcl-exmple}
  \end{minipage}%
\end{figure}

\begin{definition}[Syntax of $\pgcl$]
  The set of \emph{commands} $\pC$ and \emph{guarded commands} $\pGC$ written in $\pgcl$ is given by the grammar in \Cref{fig:pgcl-syntax},
   where $x$ is a \emph{program variable} taken from a finite set  $\pVar$,
  $\pD$ is a \emph{distribution expression} over $\pVar$, $\pB$ is an \emph{Boolean expression} over $\pVar$, $\msg$ is a \emph{message}\footnote{Think: some string. In our DPN encoding, messages will correspond to steps $(t,\beta)$.} over $\pVar$ and $\pp$ is a (rational) \emph{probability expression} over $\pVar$.
\end{definition}
Before we formalize the details, we briefly go over the intuitive meaning of each command.
 The \emph{probabilistic assignment} $\pAssign{x}{\pD}$ samples a value from $\pD$ and assigns the result to variable $x$.
The \emph{conditioning command} $\pObserve{\pB}$ checks whether $\pB$ holds and proceeds if the answer is yes. Otherwise, the current execution is discarded as if it never happened.\footnote{One can think of probabilistic programs as a stochastic simulator. Failing an observation then means that the simulation encountered an unrealistic result that should not be included in the result. A naive way to achieve this, is rejection sampling: discard the result and attempt to obtain another sample by restarting the simulation from scratch. More efficient approaches are discussed in~\cite{ppl2018}.}
The command $\pDump~\msg$ writes the value of $\msg$ to the program's log -- an append-only list of messages, which we will use to output runs.
The sequential composition $\pCompose{\pC_1}{\pC_2}$ first executes $\pC_1$, followed by $\pC_2$.
The loop $\pDood{\pGC}$ executes the guarded command $\pGC$ until no guard in $\pGC$ is enabled anymore.
More precisely, consider the loop
\[
   \pDood{\pGuard{\pB_1}{\pp_1}{\pC_1} \pChoice{} \ldots \pChoice{} \pGuard{\pB_n}{\pp_n}{\pC_n}}.
\]
If none of the guards $\pB_1, \ldots, \pB_n$ hold, the loop terminates. Otherwise, the loop randomly executes a command $\pC_i$ whose guard holds, and repeats.
The probability of executing a command is determined by the values of the expressions $\pp_1,\ldots, \pp_n$.\footnote{To ensure well-definedness, we assume that the values of $\pp_1,\ldots, \pp_n$ sum up to $\leq 1$.}
That is, if guard $\pB_i$ holds and $m$ is the total number of guards that currently hold, then the command $\pC_i$ is executed with probability $\pp_i \cdot \nicefrac{n}{m}$.
Similarly to loops, the conditional $\pIffi{\pGC}$ randomly executes one of the commands in $\pGC$ whose guard holds, but terminates afterwards.

\noindent
\emph{Example 4.}\customlabel{ex4}{4}
  Consider the $\pgcl$ program $\pC$ in \Cref{fig:pgcl-exmple}. 
  $\pC$ first randomly assigns to $x$ a value between $1$ and $3$ with probability $\nicefrac{1}{3}$ each. 
  If $x = 1$, it always assigns $4$ to $y$.
  Otherwise, it either assigns $5$ or $x + 2$ to $y$. Since only two out of three guards hold for $x > 1$, the probability of executing each assignment is $\frac{\nicefrac{1}{3}}{\nicefrac{2}{3}} = \nicefrac{1}{2}$.
   
  If, for the moment, we ignore the $\pObserve{{\color{black}B}}$, then the probability of terminating with $y = 4$ is $\nicefrac{1}{3} + \nicefrac{1}{3} \cdot \nicefrac{1}{2} = \nicefrac{1}{2}$. 
  
  Analogously, the probability to stop with $y = 5$ is $\nicefrac{2}{3} \cdot \nicefrac{1}{2} + \nicefrac{1}{3} \cdot \nicefrac{1}{2} = \nicefrac{1}{2}$.
  
  How does conditioning on ${\color{black}B}$ affect those probabilities?
  For ${\color{black}B = (y = 5)}$, we discard all executions except those that stop with $y = 5$. Hence, the probability to stop with $y = 5$ is 1, and 0 for any other value of $y$.
  For ${\color{black}B = (x > 1)}$, we consider only those executions that assign $2$ or $3$ to $x$. Both assignments happen now with probability $\nicefrac{1}{2}$. Hence, $y : = 4$ is never executed. The probability of stopping with $y = 4$ changes to $\nicefrac{1}{2} \cdot \nicefrac{1}{2} = \nicefrac{1}{4}$, \ie the probability of assigning $2$ to $x$ and executing the last assignment. Analogously, the probability of stopping with $y = 5$ changes to $\nicefrac{3}{4}$.
  For ${\color{black}B = (x = 1 \wedge y = 5)}$, there is no feasible execution. Hence, we obtain a \emph{sub}distribution that is zero for every value.
  \hfill{\footnotesize$\triangle$}

To assign formal semantics to $\pgcl$ programs, we first define program states and discuss how expressions are evaluated.
\smallskip
\noindent
\emph{States and expressions.}
The set $\pStates ~=~ \left\{~ \pState ~|~ \pState\colon \pVar \to \rats~\right\}$ of \emph{program states}  consists of all assignments of rational numbers to program variables. 

Furthermore, the set of \emph{program logs} $\pLogs ~=~ \{ \pLog ~|~ \pLog \in \textbf{Msg}^{*} \}$, where $\textbf{Msg}$ is an infinite set of \emph{messages} of interest, e.g. the set of all steps of a DPN.

We denote by $\psl = \pStates \times \pLogs$ the set of all pairs of a state $\pState$ and a program log $\pLog$.
We abstract from concrete syntax for expressions. Instead, we assume that, for every state, 
\emph{distribution expressions} $\pD$ evaluate to distributions over rationals,
\emph{Boolean expressions} $\pB$ evaluate to $\bool = \{\true,\bot\}$,
\emph{messages} $\msg$ evaluate to $\textbf{Msg}$, 
and expressions $\pp$ evaluate to rationals in $[0,1]$,
respectively. 
Formally, we assume the following evaluation functions for those expressions:
\begin{align*}
  	\pSem{\pB}\colon \pStates \to \bool
  	~~
  	\pSem{\pD}\colon \pStates \to \pDist{\rats}
  	~~
  	\pSem{\msg}\colon \pStates \to \textbf{Msg}
  	~~
  	\pSem{\pp}\colon \pStates \to [0,1]\cap\rats
\end{align*}
Throughout this paper, we use common expressions, such as $x + y < 17$, $\iverson{x > 0}$, and $\pFont{uniform(1,x+2)}$, where the corresponding evaluation functions are straightforward, e.g. 
$\lambda \pState . \pState(x) + \pState(y) < 17$, $\lambda \pState. \iverson{\pState(x) > 0}$,
and $\lambda \pState . \pFont{uniform}(1,\pState(x)+2)$.\footnote{We use lambda-notation, e.g., $\lambda \pState. 1 + \pState(x)$, to define functions that depend on $\pState$.}
\smallskip
\noindent
\emph{Semantics.}
A standard approach to assign (denotational) semantics to ordinary programs is to view them as a state transformer
$\pSem{\pC}\colon \pStates \to 2^{\pStates}$. That is, $\pC$ produces a set of output states for any given input state.
The set of output states can be empty, e.g. if the program enters an infinite loop.

For probabilistic programs, we obtain a more fine-grained view~\cite{Kozen79}, because the probability of every output state can be quantified. 
Similarly, the semantics of $\pgcl$ commands $\pC$ is a function
$
  \pSSem{\pC}\colon \psl \to \pSubDist{\psl},
$ 
that maps every initial state to a \emph{subdistribution over output states}. 
We consider \emph{sub}distributions, because we may lose probability mass, e.g. due to nontermination.

We formalize $\pSSem{\pC}$ in two steps: We first define a function $\pSem{\pC}$ that treats observation failures like nontermination, i.e. failing an observation loses probability mass. 
After that, we \emph{normalize} $\pSem{\pC}$ by redistributing the lost probability mass among the feasible executions. Formally:
\begin{definition}[Semantics of $\pgcl$]\label{def:pgc-semantics}
The (sub-)distribution $\pSSem{\pC}(\pState,\pLog)$ computed by $\pgcl$ program $C$ for initial state-log pair $(\pState,\pLog)$ is defined as
\[
  \pSSem{\pC}(\pState,\pLog) ~=~ \pNormalize{\pSem{\pC}(\pState,\pLog)} ~=~ \frac{\pSem{\pC}(\pState,\pLog)}{\sum_{(\pState',\pLog') \in \psl} \pSem{\pC}(\pState,\pLog)(\pState', \pLog')},
\]	
where the (unnormalized) transformer $\pSem{\pC}\colon \psl \to \pSubDist{\psl}$
is defined inductively on the structure of (guarded) commands in \Cref{fig:formal-pgcl-semantics}.
\end{definition}
Intuitively, $\pSSem{\pC}(\pState,\pLog)(\pState', \pLog')$ is the probability that executing $\pgcl$ program $\pC$ on initial program state $\pState$ and log $\pLog$ terminates in program state $\pState'$ with log $\pLog'$.

While our semantics is precise, we remark that there also exist ``sampling-based semantics'' for probabilistic programs in the literature, e.g.~\cite{dahlqvist20}, which guarantee that, given enough samples, the computed (sub)distribution will converge towards the subdistribution $\pSem{\pC}(\pState,\pLog)$. 
Depending on the chosen inference engine applied, we thus either get exact or approximate guarantees.

\begin{figure}[t]
\centering
\begin{tabular}{l@{\hspace{1cm}}l}
    \toprule
	  \textbf{Program $\pC$ (resp. $\pGC$)} 
	& 
	  \textbf{$\pSem{\pC}(\pState,\pLog)$ (resp. $\pSem{\pGC}(\pState,\pLog)$)} 
	  \\
	\midrule
	  $\pAssign{x}{\pD}$
	&
	  $\sum_{q \in \rats} \pSem{\pD}(\pState)(q) \cdot \pDirac{(\pState\sUpdate{x}{q}, \pLog)}$
	\\
	  $\pDump~\msg$
	&
	  $\pDirac{(\pState,\pLog\,\pSem{\msg}(\pState))}$
	\\
	  $\pObserve{\pB}$
	&
	  $\pSem{\iverson{\pB}}(\pState) \cdot \pDirac{(\pState,\pLog)}$
	\\
	  $\pCompose{\pC_1}{\pC_2}$
	&
	  $\sum_{(\pState',\pLog') \in \psl}  \pSem{\pC_1}(\pState,\pLog)(\pState',\pLog') \cdot \pSem{\pC_2}(\pState',\pLog')$
	\\
	  $\pIffi{\pGC}$
	&
	  $\frac{\pSem{\pDone{\pGC}}(\pState)}{\pSem{\pBranchProb{\pGC}}(\pState)} \cdot \pSem{\pGC}(\pState,\pLog)$
	\\
	  $\pGC_1\,\pChoice{}\,\pGC_2$
	&
	  $\pSem{\pGC_1}(\pState,\pLog) + \pSem{\pGC_2}(\pState,\pLog)$
	\\
	  $\pGuard{\pB}{\pp}{\pC_1}$
	&
	  $\pSem{\iverson{\pB}}(\pState) \cdot \pSem{\pp}(\pState) \cdot \pSem{\pC_1}(\pState,\pLog)$
	\\
	  $\pDood{\pGC}$
	&
	  $\lim_{n \to \infty} \pSem{\pC_n}(\pState,\pLog)$, where
	\\
	&
      $\pSem{\pC_0}(\pState,\pLog) ~=~ \lambda (\pState',\pLog').~0$
	\\
	&
      $\pSem{\pC_{n+1}}(\pState,\pLog) ~=~ (1-\pSem{\pDone{\pGC}}(\pState)) \cdot \pDirac{(\pState,\pLog)}$ \\
    & $\qquad\qquad\qquad\qquad + \pSem{\pCompose{\pIffi{\pGC}}{\pC_n}}(\pState,\pLog)$
	\\
	\bottomrule
	\\
\end{tabular}
\caption{Semantics of $\pgcl$ programs. 
Here, $(\pState,\pLog) \in \psl$.
We denote by $\pState\sUpdate{x}{v}$ the update of $\pState$ in which the value of $x$ is set to $v$, \ie
   $\pState\sUpdate{x}{v}(y) = v$ if $y = x$, and $\pState\sUpdate{x}{v}(y) = \pState(y)$, otherwise. $\pLog \pSem{\msg}(\pState)$ denotes the concatenation of log $\pLog$ and the evaluation of message $\msg$ in $\pState$. $\pDone{\pGC}$ and $\pBranchProb{\pGC}$ are defined further below.}
\label{fig:formal-pgcl-semantics}
\end{figure}
\smallskip
\noindent
\emph{Definition of $\pSem{\pC}$.}
We now go over the definition of $\pSem{\pC}$ in \Cref{fig:formal-pgcl-semantics} and formal notation that is not explained in the caption. In each case, we are given an initial state-log pair $(\pState,\pLog) \in \psl$ and have to produce a final subdistribution $\pSem{\pC}(\pState,\pLog) \in \pSubDist{\psl}$ depending on program $\pC$.

For the \emph{probabilistic assignment $\pAssign{x}{\pD}$}, initial state $\pState$ and log $\pLog$, we sample a rational value $q$ from the distribution $\pSem{\pD}(\pState) \in \pDist{\rats}$ and assign that value to variable $x$ -- the program state is thus updated to $\pState\sUpdate{x}{q}$.
To compute the final subdistribution, we sum over all possible samples $q \in \rats$, weighing each sample by its probability $\pSem{\pD}(\pState)(q)$, and require that the final state is the updated one using the Dirac distribution $\pDirac{(\pState\sUpdate{x}{q}, \pLog)}$.
Notice that $\pAssign{x}{\pD}$ behaves like a standard assignment if $\pSem{\pD}(\pState)$ is a Dirac distribution (see \Cref{app:det-assign}).

For the command $\pDump~\msg$, we evaluate the message $\msg$ in the current and append the result to the current program log. The final subdistribution is thus the the Dirac distribution wrt. the current program state and the updated log.

For the \emph{conditioning command} $\pObserve{\pB}$, the initial state $\pState$ and log $\pLog$ remain unchanged if $\pB$ holds in $\pState$ -- we return the Dirac distribution $\pDirac{(\pState,\pLog)}$. Otherwise, we lose all probability mass and the final subdistribution is $\lambda(\pState',\pLog').0$.\footnote{Recall that we use a separate normalization step to account for this loss.}
Formally, we denote by $\iverson{\pB}$ the indicator function of $\pB$ (\ie, $\pSem{\iverson{\pB}}(\pState) = 1$ if $\pSem{\pB}(\pState) = \true$; otherwise, $\iverson{\pB}(\pState) = 0$).
The semantics of $\pObserve{\pB}$ is then $\pSem{\iverson{\pB}}(\pState) \cdot \pDirac{(\pState,\pLog)}$.

For the \emph{sequential composition} $\pCompose{\pC_1}{\pC_2}$, we return the subdistribution obtained from running $\pC_2$ on every state-log pair $(\pState',\pLog')$ weighted by the likelihood that executing $\pC_1$ on the initial state-log pair $(\pState,\pLog)$ terminates in $(\pState',\pLog')$.

The semantics of conditions $\pIffi{\pGC}$ and loops $\pDood{\pGC}$ depends on the guarded command $\pGC$, which we consider first.
If $\pGC$ is a branch of the form $\pGuard{\pB}{\pp}{\pC}$, then $\pC$ can be executed with probability $\pp$ if $\pB$ holds. The resulting subdistribution is thus $\iverson{\pB} \cdot \pp \cdot \pSSem{\pC}(\pMu)$.
If $\pGC$ is a choice $\pGC_1 \pChoice{} \pGC_2$, we sum the subdistributions computed for $\pC_1$ and $\pC_2$.

The semantics of $\pIffi{\pGC}$ executes some branch in $\pGC$ whose guard is enabled and normalizes among the enabled branches. That is, it uniformly distributes the probabilities of branches whose guard is not enabled among the branches whose guard is enabled.
To formalize this behaviour, we use two auxiliary definitions. 
First, $\pDone{\pGC}$ counts how many guards in $\pGC$ hold:
  \begin{align*}
     \pDone{\pGC} 
     ~=~
     \begin{cases}
     	\iverson{\pB}, & \textnormal{if}~ \pGC = (\pGuard{\pB}{\pp}{\pC}) \\
     	\pDone{\pGC_1} + \pDone{\pGC_2} & \textnormal{if}~ \pGC = (\pGC_1\pChoice{}\pGC_2) \\
     \end{cases}	
  \end{align*}
Second, $\pBranchProb{\pGC}$ determines the total probability of all guards that hold:
  \begin{align*}
     \pBranchProb{\pGC}
     ~=~
     \begin{cases}
     	\iverson{\pB} \cdot \pp, & \textnormal{if}~ \pGC = (\pGuard{\pB}{\pp}{\pC})\\
     	\pBranchProb{\pGC_1} + \pBranchProb{\pGC_2} & \textnormal{if}~ \pGC = (\pGC_1\,\pChoice{}\,\pGC_2) \\
     \end{cases}	
  \end{align*}
The subdistribution obtained from executing $\pIffi{\pGC}$ on $(\pState,\pLog)$ is then given by $\pSem{\pGC}(\pState,\pLog)$ normalized by $\pBranchProb{\pGC}$. If no guard in $\pGC$ holds, \ie $\pDone{\pGC}$ evaluates to $0$, then the final subdistribution evaluates to $0$ as well.

The semantics of the loop $\pDood{\pGC}$ is defined as the limit of the distributions produced by its finite unrollings: If no loop guard holds, we terminate with probability one in the initial state $(\pState,\pLog)$ and thus return the Dirac distribution $\pDirac{(\pState,\pLog)}$; otherwise, we return the distribution obtained from executing the loop body followed by the remaining loop unrollings.\footnote{Technically, our semantics computes the least fixed point of the loop's finite unrollings $\pC_n$, which is standard when defining program semantics, see e.g. \cite{Kozen79}.}

 \section{From DPNs to $\pgcl$ Programs}
 \label{sec:encoding}
 
 We now develop a $\pgcl$ program $\pCN$ that simulates the runs of a DPN $N$  
for a given scheduler 
and a set of goal states
 such that (1) every execution of $\pCN$ corresponds to a run of $\net$ and vice versa,
 and (2) the probability distribution of $\pCN$ equals the distribution of all of the net's runs that do not visit a goal state before all of their steps have been fired. 
 We will discuss in \Cref{sec:tasks} how this distribution can be used for process mining tasks beyond simulation.

We present the construction of $\pCN$ step by step: we first discuss the setup and how we encode net states.
In \Cref{sec:encoding:construction}, we construct $\pCN$.
Finally, \Cref{sec:correcntess} addresses why the constructed probabilistic program is correct.

\subsection{Setup and Conventions}

We first consider all dependencies needed for constructing $\pCN$.
Throughout this section, we fix a DPN $\net = \tup{P, T, F, l, A, V, \Delta, \preco,\postco}$,
an initial state $(M_0,\alpha_0)$, a scheduler $\scheduler$ of $\net$, and a set of goal states $\G$.
For simplicity, we assume that all data variables evaluate to rational numbers, \ie $\Delta = \rats$, that $\G$ contains all deadlocked net states, and that membership in $\G$ for non-deadlocked states can be expressed as a Boolean formula $\isGoal$ over net states.\footnote{Examples of $\G$ include (beside deadlocked states) the set of all states, where some final marking has been reached or a variable is above some threshold.}

Furthermore, we assume that 
$P = \{p_1, \ldots, p_{\noP}\}$,
$T = \{t_1, \ldots, t_{\noT}\}$, and
$V = \{v_1, v_2, \ldots, v_{\noV}\}$ for some natural numbers $\noP,\noT,\noV \in \nats$.

We use the following program variables in our construction:
\begin{compactitem}[$\bullet$]
	\item For every place $p \in P$, $\vp{p}$ stores how many tokens are currently in $p$.
	\item For every variable $v \in V$, $\vd{v}$ stores the current value of $v$ and $\vdd{v}$ is an internal program variable used for updating the value of $\vd{v}$ when firing a transition.
	\end{compactitem}
Every underlined variable, e.g. $\vp{p}$, corresponds to a concept of the net $\net$, e.g. the tokens in the place $p$. Hence, it is straightforward to reconstruct a net state from a program state $\pState$.
More formally, the marking $\lDecode{M}{\pState}$ encoded by $\pState$ is given by $\lDecode{M}{\pState}(p) = \pState(\vp{p})$ for all places $p \in P$.
Analogously, the valuation $\lDecode{\alpha}{\pState}$ encoded by $\pState$ is given by $\lDecode{\alpha}{\pState}(v) = \pState(\vd{v})$ for all $v \in V_{\net}$.

We write $\schedulerT$ and $\schedulerV$ to refer to the transition and data component obtained from evaluating the scheduler $\scheduler$ in the current program state.
That is, if $\pState$ is the current program state, then $(\schedulerT,\schedulerV) = \scheduler(\lDecode{M}{\pState},\lDecode{\alpha}{\pState})$.
Notice that both $\schedulerT$ and $\schedulerV$ can be represented as (distribution) expressions over program variables.
Similarly, we use the Boolean expression $\isGoal$ to check whether a (non-deadlocked) state is a goal state in $\G$.

Finally, we lift our notation $\encode{\hdots}$ to expressions over $V \cup V'$. That is, we denote by $\encode{\preco(t)}$ the expression $\preco(t)$ in which every variable $v \in V$ has been replaced by $\vd{v}$.
Analogously, $\encode{\postco(t)}$ is the expression $\postco(t)$ in which every variable $v \in V$ has been replaced by $\vd{v}$ and every variable $v' \in V'$ has been replaced by $\vdd{v}$.

\subsection{Simulating Net Runs in $\pgcl$}
\label{sec:encoding:construction}

Intuitively, the $\pgcl$ program $\pCN$ simulates the runs by probabilistically selecting and firing enabled transitions in a loop until a goal state in $\G$ has been reached. In every loop iteration, will add exactly one step to the program log.

\Cref{fig:pcn} depicts how the above behaviour is implemented in the $\pgcl$ program $\pCN$, where the $\pgcl$ programs for
 setting up the initial net state ($\pC_{\mathit{init}}$), 
checking whether transition $t_i$ is enabled ($\pB_{\mathit{enabled}}(t_i)$), and firing transition $t_i$ ($\pC_{\mathit{fire}}(t_i)$) are discussed in the following.

\begin{figure}[t]
 \begin{align*}
   & \pC_{\mathit{init}}; \\ 
   & \pDo \tag*{\com{$\pLoop$: main loop fires enabled transitions until a goal state is reached}} \\
   & \pTab \pGuard{\neg\isGoal ~\wedge~ \pB_{\mathit{enabled}}(t_1)}{\schedulerT(t_1)}{\pC_{\mathit{fire}}(t_1)} \\ 
   & \pTab\pTab\vdots \\
   & \pChoice{}~ \pGuard{\neg\isGoal ~\wedge~ \pB_{\mathit{enabled}}(t_{\noT})}{\schedulerT(t_{\noT})}{\pC_{\mathit{fire}}(t_{\noT})} \\ 
   & \pOd 
 \end{align*}
	\caption{The $\pgcl$ program $\pCN$ simulating the net $\net$.}
	\label{fig:pcn}
\end{figure}

\smallskip
\noindent
\emph{$\pC_{\mathit{init}}$.}
The program $\pC_{\mathit{init}}$ below sets up the initial net state by assigning to variable $\vp{p_i}$ the number of tokens $M_0(p_i)$ initially in place $p_i$ and to variable $\vd{v_j}$ the initial value $\alpha_0(v_j)$. We also initialize $\vdd{v_j}$ with $\alpha_0(v_j)$ such that primed and unprimed variables store the same values before every loop iteration.
\begin{align*}
  \pC_{\mathit{init}}\colon\quad 
  & \pAssign{\vp{p_1}}{M_0(p_1)};~\hdots;~\pAssign{\vp{p_{\noP}}}{M_0(p_{\noP})}; \\
  & \pAssign{\vd{v_1}}{\alpha_0(v_1)};~\hdots;~\pAssign{\vd{v_{\noV}}}{\alpha_0(v_{\noV})}; \\
  & \pAssign{\vdd{v_1}}{\alpha_0(v_1)};~\hdots;~\pAssign{\vdd{v_{\noV}}}{\alpha_0(v_{\noV})};
\end{align*}
\smallskip
\noindent
\emph{$\pB_{\mathit{enabled}}(t)$.}
The guard below checks whether transition $t \in T$ can be fired:
\begin{align*}
  \pB_{\mathit{enabled}}(t)\colon\quad 
  & \encode{\preco(t)} ~\wedge~ \bigwedge_{p \in \pre{t}} \vp{p} \geq 1
\end{align*}

\smallskip
\noindent
\emph{$\pC_{\mathit{fire}}(t)$.}
For a transition $t \in T$, let $\pre{t} = \{ q_{1}, \ldots, q_{m} \} \subseteq P$ and $\post{t} = \{ r_{1}, \ldots, r_{n} \} \subseteq P$. Moreover, let $V'(\postco(t)) = \{ u'_1, \ldots, u'_k \}$ be the variables that are potentially modified by firing $t$.
Finally, we denote by $\getStep{t}$ the \emph{message} in $\textbf{Msg}$ representing a step $(t,\beta)$ of the net, where $\beta$ is given by the current values of the program's variables. Formally, $\getStep{t} = (t,\beta)$, where $\beta(u) = \vd{u}$ if $u \in V(t)$ and $\beta(u) = \vdd{u}$ if $u \in V'(t)$.
We then implement $\pC_{\mathit{fire}}(t)$ as follows:
\begin{align*}
  \pC_{\mathit{fire}}(t)\colon\quad &
  \pAssign{\vp{q_1}}{\vp{q_1}-1};~\ldots;~\pAssign{\vp{q_m}}{\vp{q_m}-1}; 
  \tag*{\com{remove tokens}} \\ &
  \pAssign{\vp{r_1}}{\vp{r_1}+1};~\ldots;~\pAssign{\vp{r_n}}{\vp{r_n}+1}; 
  \tag*{\com{add tokens}} \\ & 
  %
  %
  \pAssign{\vdd{u_1}}{\schedulerV(u_1')};~\ldots;~\pAssign{\vdd{u_k}}{\schedulerV(u_k')}; 
  \tag*{\com{sampling}} \\ &
  \pObserve{\encode{\postco(t)}}; 
  \tag*{\com{conditioning on the postcondition}} \\ &
  \pDump~\getStep{t}; 
  \tag*{\com{add the just performed step to the log}} \\ &
  \pAssign{\vd{u_1}}{\vdd{u_1}};~\ldots;~\pAssign{\vd{u_{k}}}{\vdd{u_{k}}} 
  \tag*{\com{update encoded data valuation}}
\end{align*}
The program first updates the program variables that encode the marking based on $\pre{t}$ and $\post{t}$. 
We then use the scheduler component $\schedulerV$ to sample new values for all potentially modified variables. 
We also observe $\encode{\postco(t)}$ to ensure that the sampled values satisfy the postcondition.
After that, we add the just performed step to the program log.
Finally, we update the encoded valuation $\alpha$ by assigning the values of primed variables to their unprimed counterparts.
 \subsection{Correctness}
 \label{sec:correcntess}
In this section, we show how the (sub)distribution produced by the $\pgcl$ program $\pCN$ relates to the probabilities of runs of the encoded net $\net$.

To formalize this relationship, we call a program state $\pState$ \emph{observable} iff its primed and umprimed variables store the same values, i.e. $\pState(\vd{u}) = \pState(\vdd{u})$ for all $u \in V_{\net}$.
We denote by $\sM{M}{\alpha}$ the unique observable program state given by $\sM{M}{\alpha}(p) = M(p)$ for all $p \in P$ and $\sM{M}{\alpha}(\vd{u}) = \alpha(u)$ for all $u \in V_{\net}$.

Our correctness theorem then intuitively states that running $\pLoop$ on observable states produces all legal runs of $\net$ with the same probability as $\net$:
\begin{restatable}[Correctness]{theorem}{theoremCorrectness}\label{thm:correctness}
  Let $\pLoop$ be the $\pgcl$ program constructed for net $\net$,
  goal states $\G$, and scheduler $\scheduler$ in \Cref{fig:pcn}. 
  For all states $(M,\alpha)$ of $\net$,
  \begin{align*}
    \pSSem{\pLoop}(\sM{M}{\alpha},\varepsilon)
    ~=~ \lambda (\pState,\sigma).
    \begin{cases}
    	 \mathbb{P}_{\scheduler}[M,\alpha]\left( \sigma  \,|\, \diamondsuit \mathcal{G} \right),
    	 & \text{if } \pState = \sM{M'}{\alpha'} \\
    	 & ~~\text{and}~ \sigma \in \Runs{M}{\alpha} \\
    	 & ~~\text{and}~ (M,\alpha)\goto{\sigma}(M',\alpha') \\
    	 0, & \text{otherwise}.
    \end{cases}
  \end{align*}
\end{restatable}
\noindent
For a detailed proof, we refer to \Cref{app:correctness}.

In other words, for every initial net state $(M,\alpha)$, executing the main loop of $\pCN$ on $\sM{M}{\alpha}$ produces the same distribution over runs as the encoded net $\net$ (for the same scheduler and set of goal states).

In particular, $\pInit$ always produces an observable program state corresponding to the initial net state $(M_0,\alpha_0)$, i.e. $\pSem{\pInit} = \lambda(\pState,\pLog). \pDirac{(\sM{M_0}{\alpha_0},\pLog)}$.
Hence, we have $\pSSem{\pCN}(\pState,\varepsilon) = \pSSem{\pLoop}(\sM{M_0}{\alpha_0},\varepsilon)$ regardless of the initial program state $\pState$. That is, $\pCN$ produces all legal runs of $\net$ starting in $(M_0,\alpha_0)$ with the same probability as the net for the given scheduler.

\section{Probabilistic Programming for Process Mining Tasks}
\label{sec:tasks}


So far, we outlined how to construct a probabilistic program from a DPN with a scheduler that, by \Cref{thm:correctness}, computes every DPN run with exactly the probability induced by the scheduler.
Our probabilistic program can be viewed both as a program that can be executed and as a statistical model that can be further analyzed with statistical inference engines.
In this section, we outline how one can leverage these views for Process Mining tasks.

\smallskip\noindent
\textbf{Log generation (with guarantees).}
By viewing probabilistic programs as executable programs, this use case immediately follows from the run (and, eventually, trace) generation capabilities of our approach.
Since every execution of the probabilistic program $\pCN$ yields a DPN run, it suffices to execute $\pCN$ $n$ times to generate a data set of $n$ DPN runs, which can be further projected to obtain an event log.
Notice that such projections are done at the run-to-trace level and require matching every step $(t,\beta)$ to an event (as well as well-crafted handling of silent transitions, if any). Like that, each such event carries ``activity payloads''
containing information about all the valuations of process variables from $V_N$ involved in executing non-silent activity $l(t)$ and stores new valuations for those variables that have been updated by
the post-condition of $t$.
By using PP, we get statistical guarantees on the data set (cf.~\cite{webPPL14}): the probability of the generated runs will converge to the run's probability induced by the selected scheduler. 
The same guarantees apply to the logs extracted from the sets of runs.

\smallskip\noindent
\textbf{Distribution analysis.}
By viewing PPs as statistical models, we can leverage statistical inference engines (cf.~\cite{gordon2014,webPPL14}) to analyze the distribution of DPN runs produced by our program.
Knowing this distribution is useful to identify, for example, whether certain runs are particularly (un)likely.
The true benefit, however, is that inference engines can also compute \emph{conditional} probabilities, e.g. ``what is the probability of reaching a marking \emph{given} that the data variable $x$ is at least $17.5$ and that transition $t_3$ has been fired at most twice?''. 
Technically, this can be achieved by inserting the command $\pObserve{(x > 17.5 \,\land\, \#t_3 \leq 2)}$ in our program, where $\#t_3$ is an injected variable that counts the number of transition firings. After that, we run an inference engine to compute the conditional probability distribution over DPN runs in which the above observation holds.

The same reasoning can be adopted for generating traces with \emph{rare events}.
Assume that we know from real data that a certain event rarely happens 
(e.g., a transition being executed twice or two data variables being equal).
We can focus on such rare events using conditional probabilities and add
a suitable command $\pObserve{\varphi}$ (where $\varphi$ encodes that rare event) 
in the probabilistic program. 
 A statistical inference engine will then produce the conditional probability 
distribution over only those runs in which the rare event (defined in $\varphi$)
happens, which enables further analysis, e.g. what events appear frequently if $\varphi$ holds.

\smallskip\noindent
\textbf{What-if analysis.}
Along the same vein, we can perform what-if analysis: to test a hypothesis over the given DPN model, it suffices to modify the scheduler and/or add conditioning commands $\pObserve{\varphi}$.
As in the previous case, this will produce conditional probability distributions over those runs in which described scenario happens. Notice that such an analysis does not require any modifications of the underlying DPN and requires minor adjustments to the scheduler and/or observations, which
can be easily incorporated without re-running the whole translation process discussed in Section~\ref{sec:encoding:construction}.

\section{Conclusion and Future Work}
\label{sec:conclusions}
This paper, for the first time, establishes a systematic and proven-correct connection between data Petri nets (DPNs) and probabilistic programming, with the goal of making powerful simulation and inference engines available to DPNs. Such engines can be used for a plethora of tasks such as trace generation, complex statistical analysis of DPN runs, what-if analysis.

\begin{figure}[t!]
\includegraphics[width=9cm]{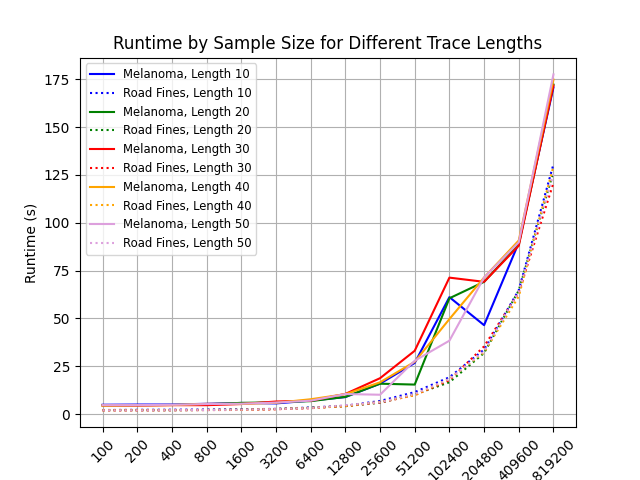}
\centering
\caption{Evaluation results for Melanoma and Road Fine, highlighting the runtime (in seconds) in relation to the number of generated runs (x-axis)}\label{fig:results}
\end{figure}

To test the feasibility of the presented approach, we developed a proof-of-concept implementation of the translation\footnote{Publicly accessible at \url{https://tinyurl.com/2hcemjkr}.} along with several examples) from PNML\footnote{\url{https://www.pnml.org}} representations of DPNs into WebPPL~\cite{webPPL14} programs that can subsequently be executed in the WebPPL environment, facilitating the simulation, analysis, and inference of the statistical model they represent.

The simulator was tested
on two nets: the Road Fine DPN taken from~\cite{MannhardtLRA16} (9 places, 19 transitions, 11 guards, 8 variables) and the Melanoma DPN (50 places, 76 transitions, 52 guards, 26 variables) taken from~\cite{Grueger21}.
For both DPNs, we used a scheduler that uniformly selects transitions and data values. The set of goal states consists of all states that are reached by runs of some fixed length.
WebPPL's MCMC inference engine was executed for various run lengths (10-50) and sample sizes (100-819200) with a 180s timeout to evaluate the performance. Figure \ref{fig:results} illustrates the runtime across five computation cycles.
Notice that the number of generated runs, depicted on the x-axis, is doubled in every step, hence the exponential increase in runtime.
Since our implementation naively follows the formal translation in \Cref{sec:encoding}, we consider the obtained runtimes encouraging and believe that there is ample space for optimizations.

For future work, we would like to investigate how probabilistic programming can be used to handle different types of monitoring tasks, where a partial execution prefix is given and has to be reproduced by the probabilistic program.  Moreover, our current work does not explicitly support silent transitions. Finally, we want to study extended simulation setups where dependencies between different runs or non-functional criteria (e.g., case arrival time or resource allocation~\cite{RosenthalTS21}) are also taken into account.

\bibliographystyle{abbrv}
\bibliography{bibliography}

\begin{thebibliography}{10}

\bibitem{mc-book}
C.~Baier and J.~Katoen.
\newblock {\em Principles of model checking}.
\newblock {MIT} Press, 2008.

\bibitem{Burattin2015}
A.~Burattin.
\newblock Plg2: Multiperspective processes randomization and simulation for
  online and offline settings.
\newblock {\em ArXiv}, abs/1506.08415, 2015.

\bibitem{Chick06}
S.~E. Chick.
\newblock Bayesian ideas and discrete event simulation: why, what and how.
\newblock In L.~F. Perrone, B.~Lawson, J.~Liu, and F.~P. Wieland, editors, {\em
  Proc. of {WSC}}, pages 96--105. {IEEE} Computer Society, 2006.

\bibitem{dahlqvist20}
F.~Dahlqvist and D.~Kozen.
\newblock Semantics of higher-order probabilistic programs with conditioning.
\newblock {\em Proc. {ACM} Program. Lang.}, 4({POPL}):57:1--57:29, 2020.

\bibitem{LeoniFM18}
M.~de~Leoni, P.~Felli, and M.~Montali.
\newblock A holistic approach for soundness verification of decision-aware
  process models.
\newblock In {\em {ER}}, volume 11157 of {\em LNCS}. Springer, 2018.

\bibitem{LeoniFM21}
M.~de~Leoni, P.~Felli, and M.~Montali.
\newblock Integrating {BPMN} and {DMN:} modeling and analysis.
\newblock {\em J. Data Semant.}, 10(1-2):165--188, 2021.

\bibitem{FelliGMRW21}
P.~Felli, A.~Gianola, M.~Montali, A.~Rivkin, and S.~Winkler.
\newblock Cocomot: Conformance checking of multi-perspective processes via
  {SMT}.
\newblock In A.~Polyvyanyy, M.~T. Wynn, A.~V. Looy, and M.~Reichert, editors,
  {\em Proc. of {BPM}}, volume 12875 of {\em LNCS}, 2021.

\bibitem{FelliGMRW22}
P.~Felli, A.~Gianola, M.~Montali, A.~Rivkin, and S.~Winkler.
\newblock Conformance checking with uncertainty via {SMT}.
\newblock In {\em Proc. of {BPM} 2022}, LNCS. Springer, 2022.

\bibitem{FelliMW22-ijcar}
P.~Felli, M.~Montali, and S.~Winkler.
\newblock Ctl\({}^{\mbox{*}}\) model checking for data-aware dynamic systems
  with arithmetic.
\newblock In J.~Blanchette, L.~Kov{\'{a}}cs, and D.~Pattinson, editors, {\em
  Proc. of {IJCAR} 2022}, volume 13385 of {\em {LNCS}}, pages 36--56. Springer,
  2022.

\bibitem{FelliMW22}
P.~Felli, M.~Montali, and S.~Winkler.
\newblock Soundness of data-aware processes with arithmetic conditions.
\newblock In X.~Franch, G.~Poels, F.~Gailly, and M.~Snoeck, editors, {\em Proc.
  of {CAiSE} 2022}, volume 13295 of {\em LNCS}, pages 389--406. Springer, 2022.

\bibitem{webPPL14}
N.~D. Goodman and A.~Stuhlm\"{u}ller.
\newblock {The Design and Implementation of Probabilistic Programming
  Languages}.
\newblock \url{http://dippl.org}, 2014.
\newblock Accessed: 2024-3-8.

\bibitem{gordon2014}
A.~D. Gordon, T.~A. Henzinger, A.~V. Nori, and S.~K. Rajamani.
\newblock Probabilistic programming.
\newblock In {\em {FOSE}}, pages 167--181. {ACM}, 2014.

\bibitem{Grueger21}
J.~Gr{\"u}ger, T.~Geyer, M.~Kuhn, S.~Braun, and R.~Bergmann.
\newblock Verifying guideline compliance in clinical treatment using
  multi-perspective conformance checking: A case study.
\newblock In {\em Process Mining Workshops}. Springer, 2022.

\bibitem{cpnbook}
K.~Jensen and L.~M. Kristensen.
\newblock {\em Coloured Petri Nets - Modelling and Validation of Concurrent
  Systems}.
\newblock Springer, 2009.

\bibitem{Kozen79}
D.~Kozen.
\newblock Semantics of probabilistic programs.
\newblock In {\em {FOCS}}. {IEEE}, 1979.

\bibitem{Law15}
A.~M. Law.
\newblock {\em Simulation Modeling \& Analysis}.
\newblock McGraw-Hill, NY, USA, 2015.

\bibitem{MannhardtLRA16}
F.~Mannhardt, M.~de~Leoni, H.~A. Reijers, and W.~M.~P. van~der Aalst.
\newblock Balanced multi-perspective checking of process conformance.
\newblock {\em Computing}, 98(4), 2016.

\bibitem{MannhardtLSL23}
F.~Mannhardt, S.~J.~J. Leemans, C.~T. Schwanen, and M.~de~Leoni.
\newblock Modelling data-aware stochastic processes - discovery and conformance
  checking.
\newblock In L.~Gomes and R.~Lorenz, editors, {\em Proc. of {PETRI} {NETS}
  2023}, LNCS. Springer, 2023.

\bibitem{Medeiros2004}
A.~Medeiros and C.~G{\"u}nther.
\newblock Process mining: Using cpn tools to create test logs for mining
  algorithms.
\newblock In {\em CPN}, 2004.

\bibitem{Mitsyuk2017}
A.~A. Mitsyuk, I.~S. Shugurov, A.~A. Kalenkova, and W.~M. {van der Aalst}.
\newblock Generating event logs for high-level process models.
\newblock {\em Simulation Modelling Practice and Theory}, 74, 2017.

\bibitem{Nakatumba2012}
J.~Nakatumba, M.~Westergaard, and W.~M. van~der Aalst.
\newblock Generating event logs with workload-dependent speeds from simulation
  models.
\newblock In {\em International Conference on Advanced Information Systems
  Engineering}. Springer, 2012.

\bibitem{PufahlWW17}
L.~Pufahl, T.~Y. Wong, and M.~Weske.
\newblock Design of an extensible {BPMN} process simulator.
\newblock In E.~Teniente and M.~Weidlich, editors, {\em Proc. of {BPM}
  Workshops}, volume 308 of {\em {LNBIP}}, pages 782--795. Springer, 2017.

\bibitem{puterman}
M.~L. Puterman.
\newblock {\em Markov Decision Processes: Discrete Stochastic Dynamic
  Programming}.
\newblock John Wiley \& Sons, Inc., USA, 1st edition, 1994.

\bibitem{RosenthalTS21}
K.~Rosenthal, B.~Ternes, and S.~Strecker.
\newblock Business process simulation on procedural graphical process models.
\newblock {\em Bus. Inf. Syst. Eng.}, 63(5):569--602, 2021.

\bibitem{ppl2018}
J.-W. van~de Meent, B.~Paige, H.~Yang, and F.~Wood.
\newblock An introduction to probabilistic programming.
\newblock {\em arXiv preprint arXiv:1809.10756}, 2018.

\bibitem{Aalst15}
W.~M.~P. van~der Aalst.
\newblock Business process simulation survival guide.
\newblock In J.~vom Brocke and M.~Rosemann, editors, {\em Handbook on Business
  Process Management 1}, pages 337--370. Springer, 2015.

\bibitem{Aalst18}
W.~M.~P. van~der Aalst.
\newblock Process mining and simulation: a match made in heaven!
\newblock In {\em SummerSim}, pages 4:1--4:12. {ACM}, 2018.

\end{thebibliography}

\appendix
\allowdisplaybreaks
\newpage
\section{Appendix}\label{app}
\subsection{Semantics of Deterministic Assignments}\label{app:det-assign}

A standard assignment $\pAssign{x}{E}$ corresponds to assigning the Dirac distribution $\pDirac{E}$ to $x$. As formalized below, we can recover the standard semantics of ordinary assignments from our semantics for probabilistic assignments.

\begin{lemma}\label{lem:det-assign}
	$\pSem{\pAssign{x}{E}}(\pState,\pLog) ~=~ \pDirac{(\pState\sUpdate{x}{\pSem{E}(\pState)},\pLog)}$.
\end{lemma}
\begin{proof}
\begin{align*}
	    & \pSem{\pAssign{x}{E}}(\pState,\pLog) \\
	~=~ & \pSem{\pAssign{x}{\pDirac{E}}}(\pState,\pLog) \tag{$\pAssign{x}{E}$ means assigning a Dirac distribution} \\
	~=~ & \sum_{q \in \rats} \pSem{\pDirac{E}}(\pState)(q) \cdot \pDirac{(\pState\sUpdate{x}{q}, \pLog)} \\
	~=~ & \sum_{q \in \rats} \iverson{q = E}(\pState) \cdot \pDirac{(\pState\sUpdate{x}{q}, \pLog)} \\  
	~=~ & \pDirac{(\pState\sUpdate{x}{\pSem{E}(\pState)},\pLog)}.  
	\tag*{$\qed$}
\end{align*}
\end{proof}
Furthermore, executing $\pC$ after a standard assignment behaves as expected. In particular, we do not explicitly have to sum over all possible final states.
\begin{lemma}\label{lem:det-seq}
	$\pSem{\pAssign{x}{E};\pC}(\pState,\pLog) ~=~ \pSem{\pC}(\pState\sUpdate{x}{\pSem{E}(\pState)},\pLog)$.
\end{lemma}
\begin{proof}
\begin{align*}
	    & \pSem{\pAssign{x}{E};\pC}(\pState,\pLog) \\
	~=~ & \sum_{(\pState',\pLog')} \pSem{\pAssign{x}{E}}(\pState,\pLog)(\pState',\pLog') \cdot \pSem{\pC}(\pState',\pLog') \\
	~=~ & \sum_{(\pState',\pLog')} \pDirac{(\pState\sUpdate{x}{\pSem{E}(\pState)},\pLog)}(\pState',\pLog') \cdot \pSem{\pC}(\pState',\pLog') \tag{\Cref{lem:det-assign}} \\
	~=~ & \pSem{\pC}(\pState\sUpdate{x}{\pSem{E}(\pState)},\pLog). \tag{evaluate Dirac distribution}\\
	\tag*{$\qed$}
\end{align*}
\end{proof}

\subsection{Proof of \Cref{thm:correctness}}\label{app:correctness}

The proof of \Cref{thm:correctness} uses a few technical, but instructive, auxiliary results.
We first present those results and defer their detailed (and somewhat tedious) proofs to the remainder of the appendix such that we can focus on \Cref{thm:correctness}.

The first lemma states that our implementation of $\pFire{t}$ correctly models firing transition $t$:
\begin{restatable}{lemma}{lemmaFire}\label{lem:fire}
  Let $\pLoop$ be the main loop constructed for net $\net$,
  goal states $\G$, and scheduler $\scheduler$. 
  If $(M,\alpha)\goto{(t,\beta)}(M',\alpha')$ then, for all $\sigma$,
  \begin{align*}
    \pSem{\pFire{t}}(\sM{M}{\alpha},\sigma)(\sM{M'}{\alpha'},\sigma\,(t,\beta))
    \quad=\quad
    \mathbb{P}_{\scheduler}[M,\alpha]\left(\beta ~|\, t \right).
  \end{align*}
  Moreover, if there is no $\beta$ or no $(M',\alpha')$ such that $\pLog = \sigma\,(t,\beta)$, $\pState = \sM{M'}{\alpha'}$, and $(M,\alpha)\goto{(t,\beta)}(M',\alpha')$, then, for all $q_1,\ldots,q_k$, then $\pSem{\pFire{t}}(\sM{M}{\alpha},\sigma)(\pState,\pLog) = 0$.
\end{restatable}
We defer the proof of \Cref{lem:fire} to \Cref{app:fire}.

The next lemma states that our implementation of the loop body $\pBody$ correctly models performing any enabled step from a given (observable) state:
\begin{restatable}{lemma}{lemmaPiffi}\label{lem:piffi}
	Let $\pDood{\pBody}$ be the main loop constructed for net $\net$, goal states $\G$, and scheduler $\scheduler$. 
	Then, for every net state $(M,\alpha)$ and every run $\sigma$, 
	\begin{align*}
	  \pSem{\pIffi{\pBody}}(\sM{M}{\alpha},\sigma) = \lambda (\pState,\pLog) .
	\begin{cases}
  	\mathbb{P}_{\scheduler}[M,\alpha]\left(t_i, \beta_i \right),
  	       & \text{if } (M,\alpha) \notin \G \\
  	       & \text{and } \pState = \sM{M_i}{\alpha_i} \\
  	       & \text{and } (M,\alpha)\goto{t_i,\beta_i}(M_i,\alpha_i) \\
  	       & \text{and } \pLog = \sigma (t_i,\beta_i)  \\
  		0, & \text{otherwise}.
  	\end{cases}
  	\end{align*}
\end{restatable}
We defer the proof of \Cref{lem:piffi} to \Cref{app:piffi}.
The next lemma collects a few properties of runs and states that we can reach by executing the main loop. In particular, this lemma shows that executions with positive probability represent legal runs of the net starting in a given initial state.
\begin{restatable}{lemma}{lemmaLike}\label{lem:like}
  Let $\pLoop$ be the main loop constructed for net $\net$,
  goal states $\G$, and scheduler $\scheduler$. 
  For every net state $(\mInit,\vInit)$, if $\pSem{\pLoop}(\sM{\mInit}{\vInit},\varepsilon)(\pState,\sigma) > 0$, then 
  \begin{enumerate}
  	\item\label{lem:like:runs} $\sigma \in \Runs{\mInit}{\vInit}$,
  	\item\label{lem:like:observable} $\pState$ is observable, and
  	\item\label{lem:like:goto} $(\mInit,\vInit)\goto{\sigma}(\lDecode{M}{\pState},\lDecode{\alpha}{\pState})$.
  \end{enumerate}
\end{restatable}

All three properties are invariants of our main loop. In particular, by construction, a step is added to the program log if and only if the loop's body is entered and a transition $t$ is fired via $\pFire{t}$.
The proof of all three properties is then by induction on the length of the given log $\sigma$ and thus also on the number of loop iterations.
We refer to \Cref{app:like} for details.
Our final lemma applies \Cref{app:piffi} to show that the main loop produces legal runs with the same probability as the net it encodes. Formally:
\begin{restatable}{lemma}{lemmaTheLoop}\label{lem:the-loop}
  Let $\pLoop$ be the main loop constructed for net $\net$, goal states $\G$, and scheduler $\scheduler$. 
  For all runs $\sigma$ and $\sigma' \in \Runs{M}{\alpha}$, 
  if $(M,\alpha)\goto{\sigma'}(M',\alpha')$ then 
  \begin{align*}
    \pSem{\pLoop}(\sM{M}{\alpha},\sigma)(\sM{M'}{\alpha'},\sigma \sigma')
    =
    \mathbb{P}_{\scheduler}[M,\alpha]\left( \sigma'  \models \diamondsuit\mathcal{G} \right).
  \end{align*}
\end{restatable}

By induction on the length of $\sigma'$. We defer the proof of \Cref{lem:the-loop} to \Cref{app:the-loop}.

\bigskip

We have now collected all auxiliary results required for proving \Cref{thm:correctness}, which we recall first:
\theoremCorrectness*
\begin{proof}
  We first consider likelihoods instead of probabilities, i.e. we show the following property -- referred to as $(\spadesuit)$ afterward.
  \begin{align*}
    \pSem{\pLoop}(\sM{M}{\alpha},\varepsilon)
    ~=~ \lambda (\pState,\sigma).
    \begin{cases}
    	 \mathbb{P}_{\scheduler}[M,\alpha]\left( \sigma  \models \diamondsuit \mathcal{G} \right),
    	 & \text{if } \pState = \sM{M'}{\alpha'} \\
    	 & ~\text{and}~ \sigma \in \Runs{M}{\alpha} \\
    	 & ~\text{and}~ (M,\alpha)\goto{\sigma}(M',\alpha') \\
    	 0, & \text{otherwise}.
    \end{cases}
  \end{align*}
  Then four cases are possible:
  \begin{enumerate}
  	\item Case: $\pState = \sM{M'}{\alpha'}$, $\sigma \in \Runs{M}{\alpha}$, and $(M,\alpha)\goto{\sigma}(M',\alpha')$. 
  	
  	  Then, by \Cref{lem:the-loop}, we have
      \begin{align*}
  	    \pSem{\pLoop}(\sM{M}{\alpha},\varepsilon)(\pState,\sigma) 
  		\quad=\quad
  		\mathbb{P}_{\scheduler}[M,\alpha]\left( \sigma  \models \diamondsuit \mathcal{G} \right).
  	  \end{align*}
  	\item Case: $\pState$ is not observable, i.e. $\pState \neq \sM{M'}{\alpha'}$ for all $(M',\alpha')$.
  	
  	  Towards a contradiction, assume $\pSem{\pLoop}(\sM{M}{\alpha},\varepsilon)(\pState,\sigma) > 0$.
  	  By \Cref{lem:like}.\ref{lem:like:observable}, $\pState$ is observable. Contradiction!
  	  Hence, $\pSem{\pLoop}(\sM{M}{\alpha},\varepsilon)(\pState,\sigma) = 0$.
  	\item Case: $\sigma \notin \Runs{M}{\alpha}$
  	
  	  	  Towards a contradiction, assume $\pSem{\pLoop}(\sM{M}{\alpha},\varepsilon)(\pState,\sigma) > 0$. By \Cref{lem:like}.\ref{lem:like:runs}, $\sigma \in \Runs{M}{\alpha}$. Contradiction! Hence, $\pSem{\pLoop}(\sM{M}{\alpha},\varepsilon)(\pState,\sigma) = 0$.
  	\item Case: $\sigma \in \Runs{M}{\alpha}$ and, for some $(M',\alpha')$, $\pState = \sM{M'}{\alpha'}$ and it does \emph{not} hold that $(M,\alpha)\goto{\sigma}(M',\alpha')$.
  	
  		  Towards a contradiction, assume $\pSem{\pLoop}(\sM{M}{\alpha},\varepsilon)(\pState,\sigma) > 0$.
  		  By \Cref{lem:like}.\ref{lem:like:runs}, $(M,\alpha)\goto{\sigma}(\lDecode{M}{\pState},\lDecode{\alpha}{\pState})$. Since $\pState = \sM{M'}{\alpha'}$, we have $(\lDecode{M}{\pState},\lDecode{\alpha}{\pState})  = (M',\alpha')$. Thus, $(M,\alpha)\goto{\sigma}(M',\alpha')$. Contradiction! Hence, $\pSem{\pLoop}(\sM{M}{\alpha},\varepsilon)(\pState,\sigma) = 0$.
  \end{enumerate}
  Put together, we conclude that the above property $(\spadesuit)$ holds.
  To complete the proof, consider the following:
  \begin{align*}
    	& \pSSem{\pLoop}(\sM{M}{\alpha},\varepsilon) \\
    ~=~ & \frac{\pSem{\pLoop}(\sM{M}{\alpha},\varepsilon)}{\sum_{(\pState,\sigma)} \pSem{\pLoop}(\sM{M}{\alpha},\varepsilon)(\pState,\sigma)} 
        \tag{\Cref{def:pgc-semantics}} \\
    ~=~ & \frac{\pSem{\pLoop}(\sM{M}{\alpha},\varepsilon)}{\sum_{\sigma' \in \textsf{Runs}(M,\alpha,\G)}\mathbb{P}_{\scheduler}[M,\alpha]\left( \sigma' \models \diamondsuit \mathcal{G}\right)} 
        \tag{apply $(\spadesuit)$ to denominator} \\
    \tag{apply $(\spadesuit)$ to numerator } \\
    ~=~ & 
    \lambda (\pState,\sigma). 
    \begin{cases}
    	 \frac{\mathbb{P}_{\scheduler}[M,\alpha]\left( \sigma  \models \diamondsuit \mathcal{G} \right)}{\sum_{\sigma' \in \textsf{Runs}(M,\alpha,\G)}\mathbb{P}_{\scheduler}[M,\alpha]\left( \sigma' \models \diamondsuit \mathcal{G}\right)},
    	 & \text{if } \pState = \sM{M'}{\alpha'} \\
    	 & ~\text{and}~ \sigma \in \Runs{M}{\alpha} \\
    	 & ~\text{and}~ (M,\alpha)\goto{\sigma}(M',\alpha') \\
    	 0, & \text{otherwise}
    \end{cases}
    \\
    \tag{by \Cref{def:run-probability}} \\
    ~=~ & 
    \lambda (\pState,\sigma). 
    \begin{cases}
    	 \mathbb{P}_{\scheduler}[M,\alpha]\left( \sigma  \,|\, \diamondsuit \mathcal{G} \right),
    	 & \text{if } \pState = \sM{M'}{\alpha'} \\
    	 & ~\text{and}~ \sigma \in \Runs{M}{\alpha} \\
    	 & ~\text{and}~ (M,\alpha)\goto{\sigma}(M',\alpha') \\
    	 0, & \text{otherwise}.
    \end{cases}
    \\
    \tag*{$\qed$}
  \end{align*}
\end{proof}


\subsection{Proof of \Cref{lem:fire}}\label{app:fire}
\lemmaFire*
\begin{proof}
Recall the definition of $\pFire{t}$, where $t \in T$, $\pre{t} = \{ q_{1}, \ldots, q_{m} \} \subseteq P$, $\post{t} = \{ r_{1}, \ldots, r_{n} \} \subseteq P$, and $V'(\postco(t)) = \{ u'_1, \ldots, u'_k \}$. We also assign names $\pC_i$ to each line for later reference:
\begin{align*}
  \pC_{1}\colon\quad &
  \pAssign{\vp{q_1}}{\vp{q_1}-1};~\ldots;~\pAssign{\vp{q_m}}{\vp{q_m}-1}; 
  \tag*{\com{remove tokens}} \\
  \pC_{2}\colon\quad &
  \pAssign{\vp{r_1}}{\vp{r_1}+1};~\ldots;~\pAssign{\vp{r_n}}{\vp{r_n}+1}; 
  \tag*{\com{add tokens}} \\ 
  \pC_{3}\colon\quad &
  \pAssign{\vdd{u_1}}{\schedulerV(u_1')};~\ldots;~\pAssign{\vdd{u_k}}{\schedulerV(u_k')}; 
  \tag*{\com{sampling}} \\
  \pC_{4}\colon\quad &
  \pObserve{\encode{\postco(t)}}; 
  \tag*{\com{conditioning on the postcondition}} \\
  \pC_{5}\colon\quad &
  \pDump~\getStep{t}; 
  \tag*{\com{add the just performed step to the log}} \\
  \pC_{6}\colon\quad &
  \pAssign{\vd{u_1}}{\vdd{u_1}};~\ldots;~\pAssign{\vd{u_{k}}}{\vdd{u_{k}}} 
  \tag*{\com{update encoded data valuation}}
\end{align*}
Then consider the following:
\begin{align*}
& \pSem{\pFire{t}}(\sM{M}{\alpha},\sigma) \\
~=~ & \pSem{\pC_{1};\pC_{2};\pC_{3};\pC_{4};\pC_{5};\pC_{6}}(\sM{M}{\alpha},\sigma) \\
~=~ & \pSem{\pC_{3};\pC_{4};\pC_{5};\pC_{6}}(\sM{M'}{\alpha},\sigma) \tag{repeatedly \Cref{lem:det-seq}, def. of $M'$} \\
~=~ & \pSem{\pAssign{\vdd{u_1}}{\schedulerV(u_1)};~\ldots;~\pAssign{\vdd{u_k}}{\schedulerV(u_k)};\pC_{4};\pC_{5};\pC_{6}}(\sM{M'}{\alpha},\sigma) \\
~=~ & \sum_{(\pState_1,\pLog_1)} \sum_{q_1 \in \rats} \schedulerV(u_1)(q_1) \cdot \pDirac{(\sM{M'}{\alpha}\sUpdate{\vdd{u_1}}{q_1}, \sigma)}(\pState_1,\pLog_1) ~\cdot \tag{\Cref{fig:formal-pgcl-semantics}} \\
    & \quad \pSem{\pAssign{\vdd{u_2}}{\schedulerV(u_2)};~\ldots;~\pAssign{\vdd{u_k}}{\schedulerV(u_k)};\pC_{4};\pC_{5};\pC_{6}}(\pState_1,\pLog_1) \\
~=~ & \sum_{(\pState_1,\pLog_1)} \sum_{q_1 \in \rats} \schedulerV(u_1)(q_1) \cdot \pDirac{(\sM{M'}{\alpha}\sUpdate{\vdd{u_1}}{q_1}, \sigma)}(\pState_1,\pLog_1) ~\cdot 
    \tag{proceed as above}\\
    & \quad \sum_{(\pState_2,\pLog_2)} \sum_{q_2 \in \rats} \schedulerV(u_2)(q_2) \cdot \pDirac{(\pState_1\sUpdate{\vdd{u_2}}{q_2}, \sigma)}(\pState_2,\pLog_2) ~\cdot~ \ldots ~\cdot~ \\
    & \qquad \sum_{(\pState_k,\pLog_k)} \sum_{q_k \in \rats} \schedulerV(u_k)(q_k) \cdot \pDirac{(\pState_{k-1}\sUpdate{\vd{u_k}}{q_k},\pLog_{k-1})}(\pState_k,\pLog_k) ~\cdot~ \\
    & \qquad\qquad \pSem{\pC_{4};\pC_{5};\pC_{6}}(\pState_k,\pLog_k) \\
~=~ & \sum_{(\pState_1,\pLog_1)} \sum_{q_1 \in \rats} \schedulerV(u_1)(q_1) \cdot \pDirac{(\sM{M'}{\alpha}\sUpdate{\vdd{u_1}}{q_1}, \sigma)}(\pState_1,\pLog_1) ~\cdot \\
    & \quad \sum_{(\pState_2,\pLog_2)} \sum_{q_2 \in \rats} \schedulerV(u_2)(q_2) \cdot \pDirac{(\pState_1\sUpdate{\vdd{u_2}}{q_2}, \sigma)}(\pState_2,\pLog_2) ~\cdot~ \ldots ~\cdot~ \\
    & \qquad \sum_{(\pState_k,\pLog_k)} \sum_{q_k \in \rats} \schedulerV(u_k)(q_k) \cdot \pDirac{(\pState_{k-1}\sUpdate{\vd{u_k}}{q_k},\pLog_{k-1})}(\pState_k,\pLog_k) ~\cdot~  \\
    & \qquad \qquad  \pSem{\iverson{\encode{\postco(t)}}}(\pState_k) \cdot \pSem{\pC_{6}}(\pState_k,\pLog_k \getStep{t}) \tag{\Cref{fig:formal-pgcl-semantics}} \\
~=~ & \sum_{(\pState_1,\pLog_1)} \sum_{q_1 \in \rats} \schedulerV(u_1)(q_1) \cdot \pDirac{(\sM{M'}{\alpha}\sUpdate{\vdd{u_1}}{q_1}, \sigma)}(\pState_1,\pLog_1) ~\cdot \\
    & \quad \sum_{(\pState_2,\pLog_2)} \sum_{q_2 \in \rats} \schedulerV(u_2)(q_2) \cdot \pDirac{(\pState_1\sUpdate{\vdd{u_2}}{q_2}, \sigma)}(\pState_2,\pLog_2) ~\cdot~ \ldots ~\cdot~ \\
    & \qquad \sum_{(\pState_k,\pLog_k)} \sum_{q_k \in \rats} \schedulerV(u_k)(q_k) \cdot \pDirac{(\pState_{k-1}\sUpdate{\vd{u_k}}{q_k},\pLog_{k-1})}(\pState_k,\pLog_k) ~\cdot~  \\
    & \qquad \qquad  \pSem{\iverson{\encode{\postco(t)}}}(\pState_k) \cdot \pSem{\pC_{6}}(\pState_k,\pLog_k \getStep{t}) \\
~=~ & \sum_{(\pState_1,\pLog_1)} \sum_{q_1 \in \rats} \schedulerV(u_1)(q_1) \cdot \pDirac{(\sM{M'}{\alpha}\sUpdate{\vdd{u_1}}{q_1}, \sigma)}(\pState_1,\pLog_1) ~\cdot \\
    & \quad \sum_{(\pState_2,\pLog_2)} \sum_{q_2 \in \rats} \schedulerV(u_2)(q_2) \cdot \pDirac{(\pState_1\sUpdate{\vdd{u_2}}{q_2}, \sigma)}(\pState_2,\pLog_2) ~\cdot~ \ldots ~\cdot~ \\
    & \qquad \sum_{(\pState_k,\pLog_k)} \sum_{q_k \in \rats} \schedulerV(u_k)(q_k) \cdot \pDirac{(\pState_{k-1}\sUpdate{\vd{u_k}}{q_k},\pLog_{k-1})}(\pState_k,\pLog_k) ~\cdot~  \\
    & \qquad \qquad  \pSem{\iverson{\encode{\postco(t)}}}(\pState_k) \cdot
      \pDirac{(\pState_k\sUpdate{\vd{u_1}}{\pState_k(\vdd{u_1})}\ldots\sUpdate{\vd{u_k}}{\pState_k(\vdd{u_k})}, \pLog_k \getStep{t})}
      \tag{repeatedly \Cref{lem:det-seq}, $u_1,\ldots,u_k$ are pairwise different}
\end{align*}
We simplify the above subdistribution by applying all Dirac distributions and inferring the possible form of $\pState_1,\ldots,\pState_k$ and $\pLog_1,\ldots,\pLog_k$:
\begin{align*}
~=~ & \sum_{q_1 \in \rats} \schedulerV(u_1)(q_1) ~\cdot \tag{simplify by applying Dirac distributions} \\
    & \quad \sum_{q_2 \in \rats} \schedulerV(u_2)(q_2) ~\cdot~ \ldots ~\cdot \sum_{q_k \in \rats} \schedulerV(u_k)(q_k) ~\cdot \\
    & \qquad \qquad  \pSem{\iverson{\encode{\postco(t)}}}(\sM{M'}{\alpha}\sUpdate{\vdd{u_1},\ldots,\vdd{u_k}}{q_1,\ldots,q_k}) ~\cdot \\
    & \qquad \qquad       
      \pDirac{(\sM{M'}{\alpha\sUpdate{u_1,\ldots,u_k}{q_1,\ldots,q_k}}, \sigma \getStep{t})}
    \\
~=~ & \sum_{q_1,\ldots,q_k \in \rats} \prod_{i=1}^{k} \schedulerV(u_i)(q_k) ~\cdot
      \tag{algebra} \\
    & \qquad \qquad  \pSem{\iverson{\encode{\postco(t)}}}(\sM{M'}{\alpha}\sUpdate{\vdd{u_1},\ldots,\vdd{u_k}}{q_1,\ldots,q_k}) ~\cdot \\
    & \qquad \qquad       
      \pDirac{(\sM{M'}{\alpha\sUpdate{u_1,\ldots,u_k}{q_1,\ldots,q_k}}, \sigma \getStep{t})}
    \\
~=~ & \mu \tag{name the above subdistribution}
\end{align*}
Now, consider $\mu(\sM{M'}{\alpha'},\sigma\,(t,\beta))$. Then the above Dirac distribution evaluates to 1 if $q_i = \beta(u_i')$ for all $i \in \{1,\ldots,k\}$ and $\getStep{t} = (t,\beta)$ (when evaluated in the current program state); otherwise, $\mu$ evaluates to $0$.
Hence,
\begin{align*}
	& \pSem{\pFire{t}}(\sM{M}{\alpha},\sigma)(\sM{M'}{\alpha'},\sigma\,(t,\beta)) \\
	~=~ & \mu(\sM{M'}{\alpha'},\sigma\,(t,\beta)) \tag{see above} \\
	~=~ & \prod_{i=1}^{k} \schedulerV(u'_i)(\beta(u_i')) ~\cdot
      \tag{see above, def. of $\encode{\postco(t)}$} \\
    & \qquad \qquad  \underbrace{\pSem{\iverson{\encode{\postco(t)}}}(\sM{M'}{\alpha}\sUpdate{\vdd{u_1},\ldots,\vdd{u_k}}{\beta(u_1'),\ldots,\beta(u_k')})}_{~=~ 1~\text{iff}~ \beta \models \postco(t)} \\
    ~=~ & \iverson{\beta \models \postco(t)} \cdot \prod_{i=1}^{k} \schedulerV(u'_i)(\beta(u_i'))
    \tag{def. of $\encode{\postco(t)}$, algebra} \\
    ~=~ & \mathbb{P}_{\scheduler}[M,\alpha]\left(\beta ~|\, t \right),
\end{align*}

Moreover, if there is no $\beta$ or no $(M',\alpha')$ such that $\pLog = \sigma\,(t,\beta)$, $\pState = \sM{M'}{\alpha'}$, and $(M,\alpha)\goto{(t,\beta)}(M',\alpha')$, then, for all $q_1,\ldots,q_k$, we have
\begin{align*}
	\pDirac{(\sM{M'}{\alpha\sUpdate{u_1,\ldots,u_k}{q_1,\ldots,q_k}}, \sigma \getStep{t})}(\pState,\pLog) ~=~ 0,
\end{align*}
i.e. $(\pState,\pLog)$ does not adequately model the net state after firing a transition, or
\begin{align*}
	\pSem{\iverson{\encode{\postco(t)}}}(\sM{M'}{\alpha}\sUpdate{\vdd{u_1},\ldots,\vdd{u_k}}{q_1,\ldots,q_k})(\pState,\pLog) ~=~ 0
\end{align*}
i.e. for no $\beta$, the postcondition of $t$ is satisfied.
Hence, by definition of $\mu$, 
\begin{align*}
	\pSem{\pFire{t}}(\sM{M}{\alpha},\sigma)(\pState,\pLog) 
	~=~ \mu(\sM{M}{\alpha},\sigma)(\pState,\pLog) ~=~ 0,
\end{align*}
which completes the proof. \qed
\end{proof}


\subsection{Proof of \Cref{lem:piffi}}\label{app:piffi}
\lemmaPiffi*
\begin{proof}
  We compute $\pSem{\pIffi{\pBody}}(\sM{M}{\alpha},\sigma)$ according to \Cref{fig:pcn,fig:formal-pgcl-semantics}:
  \begin{align*}
  	& \pSem{\pIffi{\pBody}}(\sM{M}{\alpha},\sigma) \\
  	~=~ & 
  	\frac{\pSem{\pDone{\pBody}}(\sM{M}{\alpha})}{\pSem{\pBranchProb{\pBody}}(\sM{M}{\alpha})} \cdot \pSem{\pBody}(\sM{M}{\alpha},\sigma)
  	\tag{\Cref{fig:formal-pgcl-semantics}} \\
  	~=~ & 
  	\frac{\pSem{\pDone{\pBody}}(\sM{M}{\alpha})}{\pSem{\pBranchProb{\pBody}}(\sM{M}{\alpha})} 
  	~\cdot~ 
  	\tag{\Cref{fig:pcn}} \\
  	& \quad  
  	\sum_{i = 1}^{\noT} \pSem{\pGuard{\neg\isGoal ~\wedge~ \pEnabled{t_i}}{\schedulerT(t_{i})}{\pC_{\mathit{fire}}(t_{i})}}(\sM{M}{\alpha},\sigma) 
  	\\
  	~=~ &
  	\frac{\pSem{\pDone{\pBody}}(\sM{M}{\alpha})}{\pSem{\pBranchProb{\pBody}}(\sM{M}{\alpha})} 
  	~\cdot~ \pSem{\iverson{\neg \isGoal}}(\sM{M}{\alpha}) ~\cdot~
  	\tag{\Cref{fig:formal-pgcl-semantics}} \\
  	& \quad  
  	\sum_{i = 1}^{\noT}  \pSem{\iverson{\pEnabled{t_i}}}(\sM{M}{\alpha}) \cdot \schedulerT(t_i) \cdot  \pSem{\pFire{t_i}}(\sM{M}{\alpha},\sigma) 
  	\\
  	~=~ &
  	\pSem{\iverson{\neg \isGoal}}(\sM{M}{\alpha}) ~\cdot~
  	\frac{\sum_{r = 1}^{\noT} \pSem{\iverson{\pEnabled{t_r}}}(\sM{M}{\alpha})}{\sum_{i = 1}^{\noT} \pSem{\iverson{\pEnabled{t_j}}}(\sM{M}{\alpha}) \cdot \schedulerT(t_j)} 
  	~\cdot~ 
  	\\
  	& \quad  
  	\sum_{i = 1}^{\noT}  \pSem{\iverson{\pEnabled{t_i}}}(\sM{M}{\alpha}) \cdot \schedulerT(t_i) \cdot  \pSem{\pFire{t_i}}(\sM{M}{\alpha},\sigma) 
  	\\  
  	~=~ &
  	\pSem{\iverson{\neg \isGoal}}(\sM{M}{\alpha}) ~\cdot~ \tag{algebra} \\
  	& \sum_{i = 1}^{\noT}
  	\frac{\pSem{\iverson{\pEnabled{t_i}}}(\sM{M}{\alpha}) \cdot \schedulerT(t_i) \cdot \sum_{r = 1}^{\noT} \pSem{\iverson{\pEnabled{t_r}}}(\sM{M}{\alpha})}{\sum_{j = 1}^{\noT} \pSem{\iverson{\pEnabled{t_j}}}(\sM{M}{\alpha}) \cdot \schedulerT(t_j)} 
  	\\
  	& \qquad ~\cdot~ \pSem{\pFire{t_i}}(\sM{M}{\alpha},\sigma)
  	\\
  	~=~ &
  	\pSem{\iverson{\neg \isGoal}}(\sM{M}{\alpha}) ~\cdot~ \tag{\Cref{lem:fire}} \\
  	& \sum_{i = 1}^{\noT}
  	\underbrace{\frac{\pSem{\iverson{\pEnabled{t_i}}}(\sM{M}{\alpha}) \cdot \schedulerT(t_i) \cdot \sum_{r = 1}^{\noT} \pSem{\iverson{\pEnabled{t_r}}}(\sM{M}{\alpha})}{\sum_{j = 1}^{\noT} \pSem{\iverson{\pEnabled{t_j}}}(\sM{M}{\alpha}) \cdot \schedulerT(t_j)}}_{ = \mathbb{P}_{\scheduler}[M,\alpha]\left( t_i \right)  \text{ if } (M,\alpha)\goto{t_i,\beta_i} \text{ for some } \beta_i}
  	~\cdot~ 
  	\\
  	& \qquad \lambda (\pState,\pLog). 
  	\begin{cases}
  	    \mathbb{P}_{\scheduler}[M,\alpha]\left(\beta_i ~|\, t_i \right),
  	       & \text{if } \pState = (\sM{M_i}{\alpha_i})  \\
  	       & \quad \text{and } (M,\alpha)\goto{t_i,\beta_i}(M_i,\alpha_i) \\
  	       & \quad \text{and } \pLog = \sigma (t_i,\beta_i)  \\
  		0, & \text{otherwise}
  	\end{cases}
  	\\
  	~=~ & 
  	\pSem{\iverson{\neg \isGoal}}(\sM{M}{\alpha}) ~\cdot~
  	\sum_{i = 1}^{\noT} \mathbb{P}_{\scheduler}[M,\alpha]\left( t_i \right) ~\cdot~ 
  	\tag{the factor below is $0$ whenever there is no $\beta_i$ s.t. $(M,\alpha)\goto{t_i,\beta_i}$} \\
  	& \qquad \lambda (\pState,\pLog). 
  	\begin{cases}
  	    \mathbb{P}_{\scheduler}[M,\alpha]\left(\beta_i ~|\, t_i \right),
  	       & \text{if } \pState = (\sM{M_i}{\alpha_i})  \\
  	       & \quad \text{and } (M,\alpha)\goto{t_i,\beta_i}(M_i,\alpha_i) \\
  	       & \quad \text{and } \pLog = \sigma (t_i,\beta_i)  \\
  		0, & \text{otherwise}
  	\end{cases}
  	\\
  	~=~ & 
  	\lambda (\pState,\pLog).
  	\pSem{\iverson{\neg \isGoal}}(\sM{M}{\alpha}) ~\cdot~ \\  
  	\tag{\Cref{def:step-likelihood}} \\ &
  	\sum_{i = 1}^{\noT}
  	\begin{cases}
  	    \mathbb{P}_{\scheduler}[M,\alpha]\left(t_i, \beta_i \right),
  	       & \text{if } \pState = (\sM{M_i}{\alpha_i})  \\
  	       & \quad \text{and } (M,\alpha)\goto{t_i,\beta_i}(M_i,\alpha_i) \\
  	       & \quad \text{and } \pLog = \sigma (t_i,\beta_i)  \\
  		0, & \text{otherwise}
  	\end{cases}
  	\\
  	\tag{at most one summand is $> 0$ for any given $(\pState,\pLog)$}
  	\\
  	~=~ &
  	\lambda (\pState,\pLog).  \underbrace{\pSem{\iverson{\neg \isGoal}}(\sM{M}{\alpha})}_{= 1 \text{ iff } (M,\alpha) \notin \G} \cdot 
  	\begin{cases}
  	    \mathbb{P}_{\scheduler}[M,\alpha]\left(t_i, \beta_i \right),
  	       & \text{if } \pState = (\sM{M_i}{\alpha_i})  \\
  	       & \quad \text{and } (M,\alpha)\goto{t_i,\beta_i}(M_i,\alpha_i) \\
  	       & \quad \text{and } \pLog = \sigma (t_i,\beta_i)  \\
  		0, & \text{otherwise}
  	\end{cases}
  	\\
    ~=~ & 
    \lambda (\pState,\pLog) .
	\begin{cases}
  	\mathbb{P}_{\scheduler}[M,\alpha]\left(t_i, \beta_i \right),
  	       & \text{if } (M,\alpha) \notin \G \\
  	       & \text{and } \pState = \sM{M_i}{\alpha_i} \\
  	       & \text{and } (M,\alpha)\goto{t_i,\beta_i}(M_i,\alpha_i) \\
  	       & \text{and } \pLog = \sigma (t_i,\beta_i)  \\
  		0, & \text{otherwise}.
  	\end{cases}
  \end{align*}
  The above function is equal to our claim. \qed
\end{proof}


\subsection{Proof of \Cref{lem:like}}\label{app:like}
\lemmaLike*
\begin{proof}

    All three properties are invariants of our main loop. In particular, by construction, a step is added to the program log if and only if the loop's body is entered and a transition $t$ is fired via $\pFire{t}$.
    
    We show the following slightly more general property:
    If 
    \begin{enumerate}
    	\item[(a)] $\sigma \in \Runs{\mInit}{\vInit}$,
    	\item[(b)] $(\mInit,\vInit)\goto{\sigma}(M,\alpha)$, and
    	\item[(c)] $\pSem{\pLoop}(\sM{M}{\alpha},\sigma)(\pState',\sigma \sigma') > 0$, 
    \end{enumerate}
    then 
    \begin{enumerate}
  	\item[(i)] $\sigma \sigma' \in \Runs{\mInit}{\vInit}$,
  	\item[(ii)] $\pState'$ is observable, and
  	\item[(iii)] $(M,\alpha)\goto{\sigma'}(\lDecode{M}{\pState'},\lDecode{\alpha}{\pState'})$.
  \end{enumerate}
  The claim then follows from the above for $\sigma = \varepsilon$ and $(M,\alpha) = (\mInit,\vInit)$.
  
  Since very loop iteration, i.e. every execution of $\pLoop$'s body, fires exactly one transition and thus adds exactly one step to the program log, every program execution that terminates in $(\pState',\sigma\sigma')$ when started in $(\sM{M}{\alpha},\sigma)$ performs exactly $\Length{\sigma'} = n \geq 0$ loop iterations.
  
  We show that for all $n \geq 0$, (a) - (c) implies (i) -- (iii), by induction on $n$. 
  
  \emph{Base case.} For $\Length{\sigma'} = n = 0$, we have $\sigma' = \varepsilon$.
  Then (i) follows immediate from (a) because $\sigma \sigma' = \sigma \in \Runs{\mInit}{\vInit}$.
  
  By (c), $\pLoop$ terminates without any loop iterations and thus $\pState' = \sM{M}{\alpha}$.
  Then (ii) follows immediately because $\sM{M}{\alpha}$ is observable by definition.
  
  Finally (iii) is trivial because $(M,\alpha)\goto{\varepsilon}\underbrace{(\lDecode{M}{\sM{M}{\alpha}},\lDecode{\alpha}{\sM{M}{\alpha}})}_{ = (M,\alpha)}$.
  
  \emph{Induction hypothesis.} Assume for an arbitrary, but fixed, $n \geq 0$ that properties (a) -- (c) imply properties (i) --- (iii). 
  
  \emph{Induction step.} For $\Length{\sigma'} = n+1$, the program $\pLoop$ performs at least one loop iteration, i.e. the guards of at least one of the loop's branches are satisfied by $(\sM{M}{\alpha},\sigma)$. By construction of the loop guards, this means $(M,\alpha) \notin \G$.
  
  Now, let $\sigma' = (t,\beta) \sigma''$, where $\Length{\sigma''} = n$.
  By (c), $(M,\alpha)$ satisfies $\pEnabled{t}$.
  Hence, we can execute $\pFire{t}$ with some positive likelihood to fire transition $t$.
  
  Then there exists a net state $(M'',\alpha'')$ such that $(M,\alpha)\goto{(t,\beta)}(M'',\alpha'')$.
  Clearly, $\sM{M''}{\alpha''}$ is observable and, by (a) and $(M,\alpha) \notin \G$,  $\sigma (t,\beta) \in \Runs{\mInit}{\vInit}$.
  Moreover, $(\mInit,\vInit)\goto{\sigma(t,\beta)}(M'',\alpha'')$ because, by (b), we have
  \[ (\mInit,\vInit)\goto{\sigma}(M,\alpha)\goto{(t,\beta)}(M'',\alpha''). \]
  Now, by (c), we have $\pSem{\pLoop}(\sM{M}{\alpha},\sigma)(\pState',\sigma \sigma') > 0$.
  Furthermore, by \Cref{lem:fire}, $\pSem{\pFire{t}}(\sM{M}{\alpha},\sigma)(\sM{M''}{\alpha''},\sigma\,(t,\beta)) > 0$.
  By construction of $\pLoop$, we can only add $(t,\beta)$ to the program log by executing $\pFire{t}$ before going back to the beginning of the loop.
  Consequently,
  \[ \pSem{\pLoop}(\sM{M''}{\alpha''},\sigma(t,\beta))(\pState',\sigma (t,\beta)\sigma'') > 0. \]
  Since $\Length{\sigma''} = n$ and properties (a) -- (c) hold for $(M'',\alpha'')$ and $\sigma(t,\beta)$, we can apply the induction hypothesis to conclude that:
  \begin{enumerate}
  	\item[(i)]  $\sigma (t,\beta) \sigma'' = \sigma \sigma' \in \Runs{\mInit}{\vInit}$;
  	\item[(ii)] $\pState'$ is observable; and 
  	\item[(iii)] $(M'',\alpha'')\goto{\sigma''}(\lDecode{M}{\pState'},\lDecode{\alpha}{\pState'})$ and thus also $(M,\alpha)\goto{(t,\beta)\sigma''}(\lDecode{M}{\pState'},\lDecode{\alpha}{\pState'})$.
  \end{enumerate} 
  Hence, properties (i) -- (iii) hold for runs $\sigma'$ with $\Length{\sigma'} = n+1$.
  \qed
\end{proof}


\subsection{Proof of \Cref{lem:the-loop}}\label{app:the-loop}
\lemmaTheLoop*
\begin{proof}
  Recall from \Cref{fig:pcn} the definition of 
  \begin{align*}
     \pLoop \quad=\quad 
     \pDood{\pBody}.	
  \end{align*}
  By \Cref{fig:formal-pgcl-semantics}, we have
  \begin{align*}
    \pSem{\pLoop}(\sM{M}{\alpha},\sigma)
    ~=~ &
  	\lim_{n \to \infty} \pSem{\pIter{n}}(\sM{M}{\alpha},\sigma),\quad\text{where} \\
    \pSem{\pIter{0}}(\pState,\pLog) 
    ~=~ &
    \lambda (\pState',\pLog').~0 \\
    \pSem{\pIter{n+1}}(\pState,\pLog) 
    ~=~ &
    (1-\pSem{\pDone{\pBody}}(\pState)) \cdot \pDirac{(\pState,\pLog)}
    + \pSem{\pCompose{\pIffi{\pBody}}{\pIter{n}}}(\pState,\pLog)
  \end{align*}
  We will show that, for all $\sigma'$ with $\Length{\sigma'} = n \geq 0$, if $(M,\alpha)\goto{\sigma'}(M',\alpha')$ then
  \begin{align*}
    \pSem{\pIter{n+1}}(\sM{M}{\alpha},\sigma)(\sM{M'}{\alpha'},\sigma \sigma')
    =
    \mathbb{P}_{\scheduler}[M,\alpha]\left( \sigma'  \models \mathcal{G} \right).
    \tag{$\dag$}
  \end{align*}
  
  By construction of $\pLoop$, every loop iteration adds exactly one step to the program log.
  Hence, to add a run $\sigma'$ with $\Length{\sigma'} = n$ to the program log, $\pBody$ must be executed $n+1$ times, where the additional execution corresponds to the final (failing) evaluation of the loop guards.
  Consequently, to determine the likelihood of terminating in $(\sM{M'}{\alpha'},\sigma \sigma')$, it suffices to unfold $\pLoop$ $n+1$ times. 
  Formally, if $n = \Length{\sigma'}$ then, for all $m \geq 0$, 
  \begin{align*}
  	\pSem{\pIter{n+1+m}}(\sM{M}{\alpha},\sigma)(\sM{M'}{\alpha'},\sigma \sigma')
  	\quad=\quad 
  	\pSem{\pIter{n+1}}(\sM{M}{\alpha},\sigma)(\sM{M'}{\alpha'},\sigma \sigma').
  	\tag{$\clubsuit$}
  \end{align*}
  Hence, if we manage to prove ($\dag$), it follows that
  \begin{align*}
  	 & \pSem{\pLoop}(\sM{M}{\alpha},\sigma)(\sM{M'}{\alpha'},\sigma \sigma') \\
  	~=~ & \left( \lim_{n \to \infty} \pSem{\pIter{n}}(\sM{M}{\alpha},\sigma) \right)(\sM{M'}{\alpha'},\sigma \sigma') \\
  	~=~ & \pSem{\pIter{\Length{\sigma'}+1}}(\sM{M}{\alpha},\sigma)(\sM{M'}{\alpha'},\sigma \sigma')(\sM{M'}{\alpha'},\sigma \sigma')
  	\tag{by ($\clubsuit$)} \\
  	~=~ & \mathbb{P}_{\scheduler}[M,\alpha]\left( \sigma'  \models \mathcal{G} \right). \tag{by ($\dag$)}
  \end{align*}
  To complete the proof, it thus suffices to show ($\dag$). By induction on $n \geq 0$.
  
  \emph{Base case.} For $n = \Length{\sigma'} = 0$, we have $\sigma' = \varepsilon$ and, consequently, $(M,\alpha)\goto{\sigma'}(M',\alpha')$ implies $M' = M$ and $\alpha' = \alpha$.
  We distinguish two cases:
  \begin{enumerate}
  	\item $(M,\alpha) \in \G$. Then,
  	\begin{align*}
  	& \pSem{\pIter{1}}(\sM{M}{\alpha},\sigma)(\sM{M}{\alpha},\sigma) \\ 
    ~=~ &
    (1-\underbrace{\pSem{\pDone{\pBody}}(\sM{M}{\alpha})}_{ = 0 \text{ since } \sM{M}{\alpha} \not\models \neg\isGoal}) \cdot \underbrace{\pDirac{(\sM{M}{\alpha},\sigma)}(\sM{M}{\alpha},\sigma)}_{ = 1} \\
    &
    + \underbrace{\pSem{\pCompose{\pIffi{\pBody}}{\pIter{0}}}(\pState,\pLog)	(\sM{M}{\alpha},\sigma)(\sM{M}{\alpha},\sigma)}_{ = 0 \text{ by the semantics of $;$ and $\pIter{0}$ } } \\
    ~=~ & 1 \\
    ~=~ & \mathbb{P}_{\scheduler}[M,\alpha]\left( \sigma'  \models \mathcal{G} \right). \tag{by \Cref{def:run-likelihood}}
  	\end{align*}

  	\item $(M,\alpha) \notin \G$. Then, since we assume that all deadlocked state are in $\G$,
  	\begin{align*}
  	& \pSem{\pIter{1}}(\sM{M}{\alpha},\sigma)(\sM{M}{\alpha},\sigma) \\ 
    ~=~ &
    (1-\underbrace{\pSem{\pDone{\pBody}}(\sM{M}{\alpha})}_{ = 1 \text{ since } \sM{M}{\alpha} \models \neg\isGoal \text{ and some $t$ is enabled}}) \cdot \underbrace{\pDirac{(\sM{M}{\alpha},\sigma)}(\sM{M}{\alpha},\sigma)}_{ = 1} \\
    &
    + \underbrace{\pSem{\pCompose{\pIffi{\pBody}}{\pIter{0}}}(\pState,\pLog)	(\sM{M}{\alpha},\sigma)(\sM{M}{\alpha},\sigma)}_{ = 0 \text{ by the semantics of $;$ and $\pIter{0}$ } } \\
    ~=~ & 0 \\
    ~=~ & \mathbb{P}_{\scheduler}[M,\alpha]\left( \sigma'  \models \mathcal{G} \right). \tag{by \Cref{def:run-likelihood}}
  	\end{align*}
  \end{enumerate}
  In both cases, $(\dag)$ holds for $n = 0$.
  
  \emph{Induction hypothesis.} Assume that our claim $(\dag)$ holds for some arbitrary, but fixed, natural nuber $n \geq 0$.
  
  \emph{Induction step.} For $\Length{\sigma'} = n + 1$, assume that $(M,\alpha)\goto{\sigma'}(M',\alpha')$ and $\sigma' \in \Runs{M}{\alpha}$; otherwise, there is nothing to show.
  Then, there exists a step $(t,\beta)$, a run $\sigma''$, and a state $(M'',\alpha'')$ such that
  \begin{enumerate}
  	\item $\sigma' = (t,\beta) \sigma''$,
  	\item $(M'',\alpha'') \notin \G$, 
  	\item $(M,\alpha)\goto{t,\beta}(M'',\alpha'')\goto{\sigma''}(M',\alpha')$, and
  	\item $\sigma'' \in \Runs{M''}{\alpha''}$.
  \end{enumerate} 
  Then, consider the following:
  \begin{align*}
    & \pSem{\pIter{n+2}}(\sM{M}{\alpha},\sigma)(\sM{M'}{\alpha'},\sigma (t,\beta) \sigma'') \\
    ~=~ & 
          (1-\underbrace{\pSem{\pDone{\pBody}}(\sM{M}{\alpha})}_{~=~1 \text{ by (2)}}) \cdot \pDirac{(\sM{M}{\alpha},\sigma)}(\sM{M'}{\alpha'},\sigma (t,\beta) \sigma'') \\
        & + \pSem{\pCompose{\pIffi{\pBody}}{\pIter{n+1}}}(\sM{M}{\alpha},\sigma)(\sM{M'}{\alpha'},\sigma (t,\beta) \sigma'')
    \\
    ~=~ & 
          \pSem{\pCompose{\pIffi{\pBody}}{\pIter{n+1}}}(\sM{M}{\alpha},\sigma)(\sM{M'}{\alpha'},\sigma (t,\beta) \sigma'') \tag{$(1 - 1) \cdot \ldots = 0$}
    \\
    ~=~ &
          \sum_{(\pState,\pLog)} \pSem{\pIffi{\pBody}}(\sM{M}{\alpha},\sigma)(\pState,\pLog)
          \cdot \pSem{\pIter{n+1}}(\pState,\pLog)(\sM{M'}{\alpha'},\sigma (t,\beta)\sigma'')
          \tag{\Cref{fig:formal-pgcl-semantics}}
    \end{align*}
    Now, notice that program logs can only grow and, by construction, $\pBody$ adds exactly one step step to the program logs (since some loop guards are satisfied as shown above). For every $\pLog$ with $\pLog \neq \sigma (t,\beta)$, $\pLog$ is then \emph{not} a prefix of the final program log $\sigma (t,\beta)\sigma''$ and, consequently, 
    $\pSem{\pIter{n+1}}(\pState,\pLog)(\sM{M'}{\alpha'},\sigma (t,\beta)\sigma'') = 0$.
    
    Furthermore, for $\pLog = \sigma (t,\beta)$, we know by \Cref{lem:piffi} that all of the above summands for $\pState \neq \sM{M''}{\alpha''}$ evaluate to $0$ as well.
    Hence, we can simplify the above sum as follows:
    \begin{align*}
    ~=~ &
          \pSem{\pIffi{\pBody}}(\sM{M}{\alpha},\sigma)(\sM{M''}{\alpha''},\sigma (t,\beta)) ~\cdot
          \\ 
        & \quad \pSem{\pIter{n+1}}(\pState,\pLog)(\sM{M'}{\alpha'},\sigma (t,\beta)\sigma'')
    \\
    ~=~ & \mathbb{P}_{\scheduler}[M,\alpha]\left( t,\beta \right) \cdot 
          \pSem{\pIter{n+1}}(\pState,\pLog)(\sM{M'}{\alpha'},\sigma (t,\beta)\sigma'') 
          \tag{by \Cref{lem:piffi}} \\
    ~=~ & \mathbb{P}_{\scheduler}[M,\alpha]\left( t,\beta \right) \cdot 
           \mathbb{P}_{\scheduler}[M'',\alpha'']\left( \sigma''  \models \mathcal{G} \right)
           \tag{by I.H., possible due to (3), (4)} \\
    ~=~ & \mathbb{P}_{\scheduler}[M,\alpha]\left( (t,\beta)\sigma''  \models \mathcal{G} \right) 
         \tag{\Cref{def:run-likelihood} and (3)} \\
    ~=~ & \mathbb{P}_{\scheduler}[M,\alpha]\left( \sigma'  \models \mathcal{G} \right).
         \tag{by (1)}
  \end{align*}
  Put together, we conclude $\pSem{\pIter{n+2}}(\sM{M}{\alpha},\sigma)(\sM{M'}{\alpha'},\sigma\sigma') = \mathbb{P}_{\scheduler}[M,\alpha]\left( \sigma'  \models \mathcal{G} \right)$, which completes the proof. \qed
\end{proof}



\end{document}